\documentclass[%
 reprint,
 superscriptaddress,
 amsmath,amssymb,
 aps,
prl,
]{revtex4-2}

\usepackage{lipsum}
\usepackage{graphicx}
\usepackage{dcolumn}
\usepackage{bm}
\usepackage{hyperref}
\usepackage{xcolor}

\newcommand*\dd{\mathop{}\!\mathrm{d}}

\begin{document}

\preprint{APS/123-QED}

\title{Nonlocal decoding of positional and correlational information during development}

\author{Alex Chen Yi Zhang}
 \email{alexchenyi.zhang@ist.ac.at}
\affiliation{%
  Institute of Science and Technology Austria, Klosterneuburg AT-3400, Austria
}%
\author{Pablo Mateu Hoyos}
\affiliation{%
  Institute of Science and Technology Austria, Klosterneuburg AT-3400, Austria
}%
\author{David Br\"uckner}
 \email{d.b.brueckner@unibas.ch}
\affiliation{%
  Biozentrum and Department of Physics, University of Basel, Basel CH-4056, Switzerland
}%
\author{Ga\v{s}per Tka\v{c}ik}
\email{gtkacik@ist.ac.at}
\affiliation{%
  Institute of Science and Technology Austria, Klosterneuburg AT-3400, Austria
}%

\date{\today}


\begin{abstract} 
In many developmental systems, cells differentiate into a tissue by reading out morphogen concentration fields, a process fundamentally limited by noise. How much can the precision of this process be improved by nonlocal information, e.g., via cell-cell communication? Using a Bayes-optimal framework, we show that positional inference depends crucially on morphogen spatial correlations and on the ``structural prior'' that encodes the geometry of the cellular lattice performing the readout. We derive upper bounds on positional information gain due to nonlocal readout and identify signal processing algorithms that approximate optimal positional inference, as well as simple chemical reaction schemes which implement such algorithms. Our theory suggests that correlational information can be exploited to significantly enhance developmental precision.
\end{abstract}

\maketitle

During organismal development, genetically identical cells reproducibly differentiate into distinct cell fates arranged into a precise spatial pattern -- a body plan. In many cases, this patterning relies on \emph{positional information}~\cite{wolpert1969positional, rogers_Morphogen_2011, kicheva_Control_2023}, a chemical ``global positioning system'' in which cells locally interpret graded morphogen concentration profiles to determine their positions within an embryo. The precision of this system is limited by the intrinsic stochasticity of the morphogen gradient and the molecular processes reading it out, as well as by the extrinsic -- i.e., environmental or embryo-to-embryo -- variability.

Although introduced more than 50 years ago, the positional information concept has been mathematically formalized relatively recently as the mutual information, $\mathrm{PI}=I(X;G)$, between {\it position} $x$ and a (possibly multi-dimensional) {\it morphogen value} $g$~\cite{dubuis_Positional_2013,tkavcik2021many}. The higher the $\mathrm{PI}$, the more accurately local morphogen levels can be mapped to positions along an embryonic or a tissue axis. In this view, $I$ bits of positional information would suffice to uniquely specify $\sim 2^I$ cell fates~\cite{bauer2021trading}.

A complementary view stipulates that the cells perform ``decoding,'' or a statistical inference, of positions from the local morphogen concentrations~\cite{hironaka_Encoding_2012, morishita_Coding_2011}. If this process were nearly Bayes-optimal (as suggested empirically~\cite{zagorski_Decoding_2017,petkova_Optimal_2019}), position could be computed as a maximum \emph{a posteriori} estimate $\hat{x}=\text{argmax}_{x^{\ast}}P(x^{\ast}|g)$ that maximizes the posterior over implied positions $x^{\ast}$. Data Processing Inequality asserts that the mutual information between true positions and \emph{any} positional estimate obtained from local morphogen readouts constitutes a lower bound on positional information, which in practice can be quite tight for optimal decoding, so that $\mathrm{PI}\simeq I(X,\hat{X})$~\cite{tkacik_Positional_2015,cepeda2019estimating,hledik2019tight}. This relationship connects the encoding and decoding views of PI.

\begin{figure}[bt]
    \centering
    \includegraphics[width=.95\linewidth]{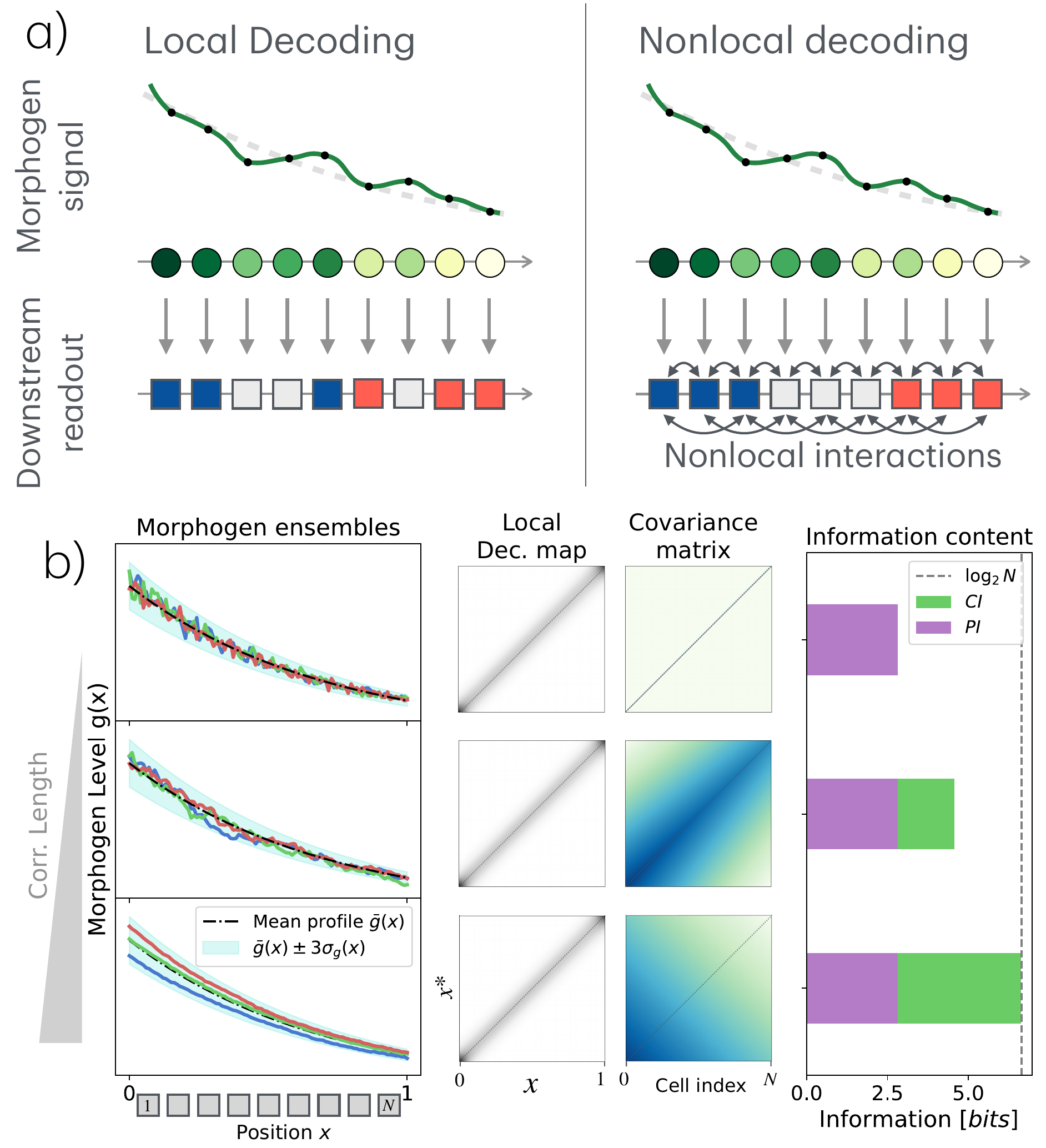}
    \caption{{\bf (a)} Cells arranged into an approximate 1D lattice (squares; here and throughout the paper, we consider $N=100$ cells) read out a local, noisy morphogen signal (circles with different shades of green) to infer their positions and commit to corresponding discrete cell fates (here, blue/white/red square colors). Signal noise can result in imprecise fate assignments (left); its impact could be mitigated by information exchange between cells, e.g., via diffusion-mediated averaging or other cell-cell coupling mechanisms (right).  {\bf (b)} Three example morphogen ensembles (first column; color lines = individual profiles; black = mean) carrying the same PI and thus having identical local decoding maps (second column)~\cite{petkova_Optimal_2019}, but differing in the spatial correlation structure (third column--see Section~S2.A in~\cite{SMnote} for details on the covariance matrix parametrization) and therefore in the amount of correlational information, CI (fourth column; dashed vertical line at $I=\log_2(N)\approx 6.6$ bits = perfect specification of every cell in the lattice.)}
    \label{fig:fig_1}
\end{figure}

Both views assume that cells make independent inferences about their own position by relying only on local morphogen readouts. This is equivalent to neglecting spatially correlated fluctuations in the morphogen signal (which could be strong in case of dominant extrinsic noise). The assumption also precludes strategies where a cell could access nonlocal morphogen readouts taken at locations of other, possibly neighboring, cells -- for instance, via cell-cell communication~\cite{piddini2009interpretation,hillenbrand2016beyond,mcgough_Finding_2024, erez_NeighborhoodInformed_2025} or by diffusion-mediated averaging~\cite{erdmann2009role,sokolowski2015optimizing}.
Recent analyses of the anterior-posterior patterning in the fruit fly {\it Drosophila} demonstrate that information beyond local PI can be contained in, and extracted from, the spatial correlations in morphogen fluctuations (Fig.~\ref{fig:fig_1}a)~\cite{mcgough_Finding_2024, erez_NeighborhoodInformed_2025}. This possibility was theoretically foreshadowed by the extension of the positional information formalism to self-organized patterning~\cite{bruckner_Information_2024}, which identified \emph{correlational information}, CI, as a measure of the additional information content of morphogen profiles due to spatial correlations in their fluctuations. Thus, even while morphogen profile ensembles with identical mean and local noise variance share the same PI, they can differ in their CI (Fig.~\ref{fig:fig_1}b).
%

Here, we develop a mathematical framework to quantify the amount of information useful for patterning contained in spatial correlations of the morphogen signal, and to predict how this information can be optimally exploited. We organize our results using Marr’s three levels of analysis of information processing systems, a paradigm well established in computational neuroscience~\cite{marr1976understanding}, and recently extended to developmental patterning~\cite{bruckner2025marrslevelsembryonicdevelopment}: we first formally state the computational problem that the cells should be solving (Level I); next, we propose signal processing algorithms that approximate optimal solutions to this computational problem (Level II); finally, we provide a proof-of-principle, minimal biochemical network that can implement such algorithms (Level III). 

\textbf{Level I: Formalizing the problem.}
At the first level, our goal is to develop a decoding framework to predict optimal bounds on the PI that a cell can gain by using nonlocal, in addition to local, morphogen information, and to determine how these bounds depend on the statistical structure of the morphogen ensemble (Fig.~\ref{fig:fig_1}b). In contrast to local decoding, where a cell $i$ relies only on its own readout $g_i$, nonlocal decoding allows a cell to infer its position from morphogen readouts $\mathbf{g}$ taken at all cellular locations $\mathbf{x}=(x_1,\dots,x_N)$. For the cell $i$, this amounts to $\hat{x}_i = \text{argmax}_{x^{\ast}_i} P(x^{\ast}_i|\mathbf{g})$; if a cell only had access to information from a limited neighborhood $\mathcal{N}_i$, the corresponding posterior would be $P(x^{\ast}_i|\{g_j\}_{j \in \mathcal{N}_i})$.
In general, the posterior can be written as

\begin{equation}
P(x^{\ast}_i|\mathbf{g})  \propto P(x^{\ast}_i) \int \dd \mathbf{x}^{\ast}_{\neg i}\;P(\mathbf{g}|\mathbf{x}^{\ast})P(\mathbf{x}^{\ast}_{\neg i}|x^{\ast}_i), 
\label{eq:general_posterior}
\end{equation}
where $\mathbf{x}^{\ast}_{\neg i}$ denotes the vector of implied positions of all cells other than $i$, and $P(\mathbf{g}|\mathbf{x}^{\ast})$ is the joint likelihood of morphogen readouts given the full (assumed) cell-position vector, $\mathbf{x}^{\ast}$.
Thus, two key factors determine the limits to nonlocal positional decoding: \emph{(i)}, the correlation structure of morphogen fluctuations as captured by $P(\mathbf{g}|\mathbf{x}^{\ast})$; and \emph{(ii)}, the ``structural prior,'' $P(\mathbf{x}^{\ast}_{\neg i}|x^{\ast}_i)$, which formalizes what can be assumed about positions of neighboring cells ($j\neq i$) that contribute their nonlocal morphogen readouts to the processing in cell $i$. The structural prior is not just a technicality: it is an essential quantity that could be empirically estimated (e.g., in the fruit fly embryo~\cite{deneke2019self}). By analyzing ensembles of embryos or tissues with measured cellular (or nuclear) ``true'' readout locations $x_i$, we could in principle estimate their joint distribution, $P(\mathbf{x})$, and consequently extract the structural prior that most closely mimics the true distribution via marginalization. The structural prior captures how reproducibly the cellular lattice geometry itself can be set up by the developmental process under study. An intriguing biological implication of optimal inference in Eq.~\eqref{eq:general_posterior} is that morphogen interpretation mechanisms would have evolutionarily adapted to the statistical regularities of the cellular lattice. 

Typically, the integral over the structural prior in Eq.~(\ref{eq:general_posterior}) will be analytically intractable. To make progress and simplify the calculations, we analyze two special cases that represent limiting forms of nonlocal decoding, as we detail below. For each case, we compute the mutual information between the true and inferred positions, $I(X,\hat{X}_{\text{nl}})$, where ``$\text{nl}$'' stands for decoding with access to nonlocal information; and compare it to the local decoding baseline, $I(X,\hat{X}_{\text{l}})$. If posteriors are unimodal and the small noise approximation holds~\cite{tkacik_Positional_2015}, positional information can be expressed as:
\begin{equation}
    \mathrm{PI}\simeq I(X,\hat{X})  \simeq - \Big\langle\log_2 {\frac{\sqrt{2\pi e \sigma_x^2(x)}}{L}} \Big\rangle_x ,
\label{eq:PI_and_poserrs}
\end{equation}
where $L$ is the embryo length and the {\it positional error}, $\sigma_x^2(x) = \langle(x-x^{\ast})^2\rangle_{P(x^{\ast}|x)}$, is the mean squared error of the inferred position $\hat{x}$ given morphogen readouts at the true position $x$, as visualized by the ``decoding map'' $P(x^{\ast}|x)$ (Fig.~\ref{fig:fig_1}b)~\cite{petkova_Optimal_2019}.
We define
\begin{equation}
    \Delta \mathrm{PI} := I(X,\hat{X}_{\text{nl}}) - I(X,\hat{X}_{\text{l}})
\label{eq:def_deltaPI}
\end{equation}
as the gain in positional information, in bits, due to the optimal utilization of nonlocal information in the form of morphogen readouts at additional positions $j \in \mathcal{N}_i$, which we assume can be communicated in an error-free fashion to position $i$. In reality, any such communication would incur extra transmission noise, limiting its utility~\cite{mugler_Limits_2016}, a constraint we neglect here to compute optimal bounds, yet essential for a more complete future theory. With this caveat in mind, we proceed below to show that $\Delta \mathrm{PI}$ strongly depends  on the form of the structural prior, by focusing on two ideal limits: the Relative (RLP) and Absolute Locations Prior (ALP), respectively.

\begin{figure}
    \centering
    \includegraphics[width=.95\linewidth]{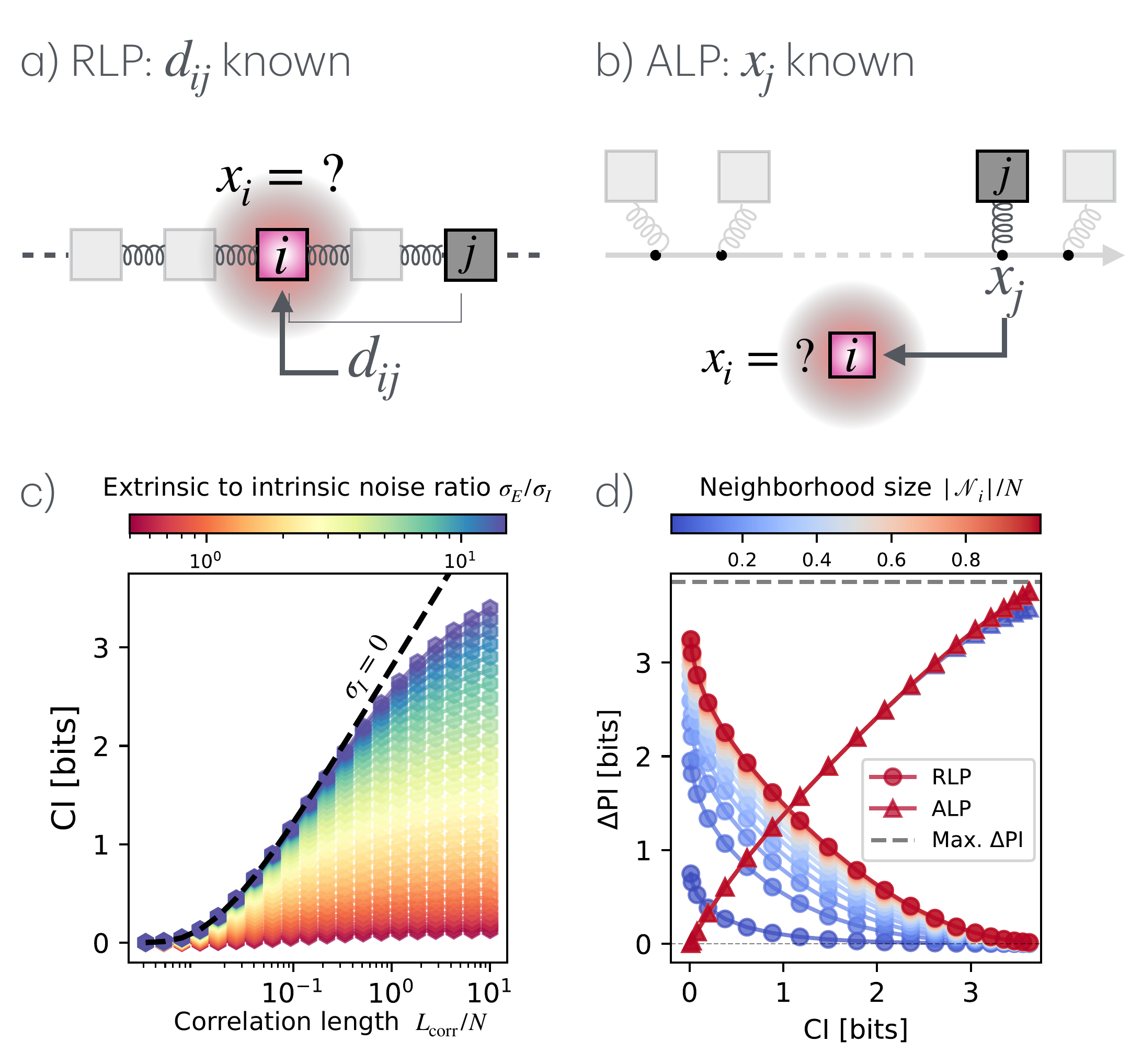}
    \caption{A focal cell $i$ (pink square) improves its positional inference with nonlocal morphogen information $g_j$ read out by neighboring cells $j\in\mathcal{N}_i$ (squares left and right of the pink focal cell). {\bf (a)} For RLP, the nonlocal signal is interpreted in the context of perfectly known relative location, $d_{ij}$, of each ``sender cell'' that conveys the nonlocal signal. In reality, the relative distances would be imperfectly known (illustrated by springs). {\bf (b)} For ALP, absolute locations, $x_j$, of ``sender cells'' are assumed known. In reality, the absolute locations would be imperfectly known (illustrated by springs).  {\bf (c)} Correlational information (CI) in the morphogen ensemble increases with the correlation length of the Gaussian fluctuations, $L_{\rm corr}$, and their magnitude $\sigma_E$ relative to the intrinsic (uncorrelated) noise with magnitude $\sigma_I$ (colorbar). {\bf (d)} Maximal PI increase with nonlocal decoding using RLP (circles) or ALP (triangles). Color indicates the neighborhood size $|\mathcal{N}_i|$ (dark blue: $\pm 1$ nearest neighbor; dark red: entire system of $N$ cells). Here and for the rest of the paper, we consider $\sigma_{E} /\sigma_{I} = 15$ (violet line in (c)). Dashed grey line: $\Delta \mathrm{PI}$ corresponding to perfect cellular specification.} 
    \label{fig:typeIandII}
\end{figure}
For the limiting case of the {\it Relative Locations Prior (RLP)}, cell $i$ decodes the morphogen by relying on a perfect knowledge about the relative positions of cell $i$'s neighbors from which nonlocal morphogen readouts are received. The corresponding structural prior is
\begin{equation}
    P_{\text{RLP}}(\mathbf{x}^{\ast}_{\neg i}|x^{\ast}_i) = \prod_{j \in \mathcal{N}_i} \delta (x^{\ast}_j - (x^{\ast}_i+d_{ij})),
\end{equation}
where $d_{ij} = x_j - x_i$ are assumed to correspond to the true values. RLP used in conjunction with Eq.~\eqref{eq:general_posterior} yields inferred positions $\hat{x}_i$ and the  corresponding positional errors (see SM~\cite{SMnote} Sec.~2.C):
%
\begin{equation}
    \sigma_{x,\text{RLP}}^2(x_i) \simeq \left ( \sum_{j,k \in \mathcal{N}_i} \bar{g}'(x_j) [C_{\mathcal{N}_i}^{-1}]_{jk} \bar{g}'(x_k) \right)^{-1}.
    \label{eq:rlp_poserrs}
\end{equation}
Here, $\bar{g}'(x)$ denotes the spatial derivative of the mean profile, and $[C_{\mathcal{N}_i}^{-1}]_{jk}$ is the $(j,k)$ element of the inverse covariance matrix of morphogen fluctuations restricted to the neighborhood $\mathcal{N}_i$.
This expression mirrors the well-established result for the positional error of local decoding from one,
\begin{equation}
\sigma^2_x(x_i) \simeq \Big(\bar{g}'(x_i)^2 / C_{ii}\Big)^{-1}, \label{stddec}
\end{equation}
or from multiple gradients~\cite{tkacik_Positional_2015, petkova_Optimal_2019}; recently, it has been applied to data in the spatial context by analogy to the established result for the multi-gradient case~\cite{erez_NeighborhoodInformed_2025}.

To evaluate the performance of this decoding strategy, we generate morphogen ensembles with identical PI but different CI by varying two key parameters: the relative amplitude of intrinsic (uncorrelated) and extrinsic (correlated) morphogen fluctuations ($\sigma_I/\sigma_E$) and the correlation length of the extrinsic fluctuations ($L_{\rm corr}/N$; Fig.~\ref{fig:typeIandII}c). We then use Eqs.~\eqref{eq:PI_and_poserrs}, \eqref{eq:def_deltaPI} to compute $\Delta \mathrm{PI}$. $\Delta \mathrm{PI}$ grows monotonically with the neighborhood size $|\mathcal{N}_i|$, with largest gains seen at short morphogen profile correlation lengths (i.e., at low CI) (Fig.~\ref{fig:typeIandII}d). Longer range correlations  reduce, but do not negate, the benefits of nonlocal decoding using the RLP (SM, Fig.~S1).

For the limiting case of the {\it Absolute Locations Prior (ALP)}, cell $i$ decodes the morphogen by relying on the true positions, $x_j$, of its neighbors, so that
%
\begin{equation}
    P_{\text{ALP}}(\mathbf{x}^{\ast}_{\neg i}|x^{\ast}_i) = P(\mathbf{x}^{\ast}_{\neg i})= \prod_{j \in \mathcal{N}_i} \delta (x^{\ast}_j - x_j).
\end{equation}
Here, the positional errors read:
\begin{equation}
    \sigma_{x,\text{ALP}}^2(x_i) \simeq \Bigg( \bar{g}'(x_i)^2[C_{\mathcal{N}_i}^{-1}]_{ii} \Bigg)^{-1},
    \label{eq:alp_poserrs}
\end{equation}
leading to a compact expression for $\Delta PI$ (see SM~\cite{SMnote} Sec.~S2B):
\begin{equation}
    \Delta \mathrm{PI}_{\text{ALP}} = \frac{1}{2N} \sum_{i=1}^N[ \log_2 [C_{\mathcal{N}_i}]_{ii} + \log_2 [C_{\mathcal{N}_i}^{-1}]_{ii}.
    \label{eq:deltaPI_alp}
\end{equation}
Decoding with the ALP behaves differently from decoding with the RLP, highlighting the key role of structural priors: for ALP, performance improves with stronger correlations in the morphogen profiles (high CI) and decreases as the correlations vanish (Fig.~\ref{fig:typeIandII}d), in contrast to RLP. We will provide clearer intuition for this behavior later in the text.

The RLP and ALP define idealized bounds, each postulating a delta-function structural prior that amounts to a perfect knowledge of some aspect of the cellular lattice geometry. Real embryos, tissues, and cells are unlikely to correspond to such idealizations: variability in cell spacing and noise in intracellular communication will inevitably broaden the priors and add noise to the likelihood function. Nevertheless, as we show in the following section, these idealized bounds inform plausible and useful algorithms that can approximate positional decoding well also in more realistic situations.

\textbf{Level II: Algorithmic approximations.}
While Eqs.~\eqref{eq:rlp_poserrs},\eqref{eq:alp_poserrs}  provide a principled way to compute the available PI gain, such results are not directly operational: they specify bounds, but not how a biological system might approach them. We therefore shift our attention from the problem that the cells should be solving to the computations that they could actually instantiate to approximate the optimal solutions. This is a Level II question in Marr's paradigm. Guided by the contrasting behaviors of RLP and ALP decoding in Fig.~\ref{fig:typeIandII}, we focus on two limiting cases, one for RLP and one for ALP, so as to extract representative algorithmic strategies.
\begin{figure*}
    \centering
    \includegraphics[width=1\linewidth]{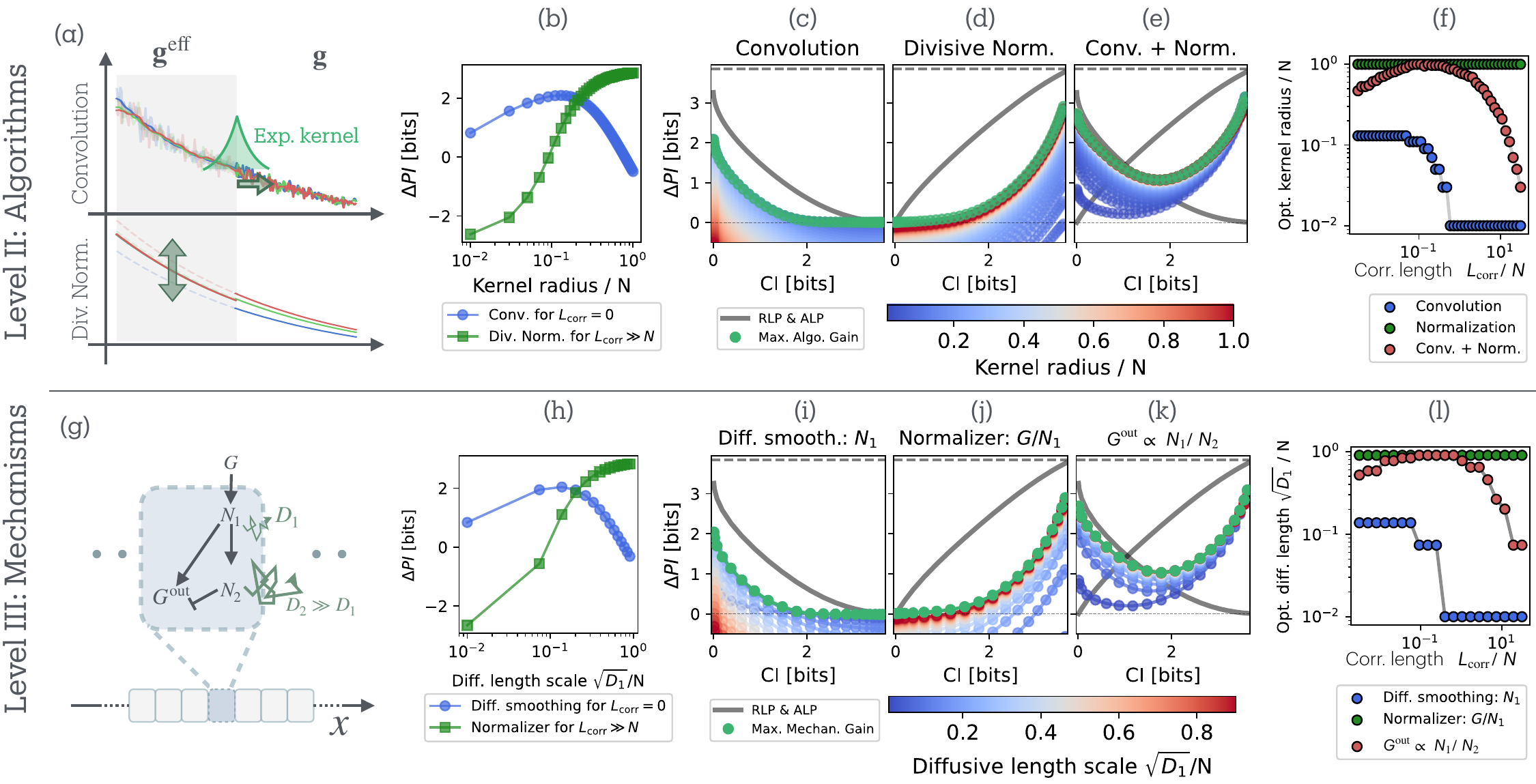}
    \caption{{\bf (a)} Algorithmic transformations map morphogen profiles $\mathbf{g}$ into effective profiles $\mathbf{g}^{\rm eff}$ via spatial convolution and/or divisive normalization; these effective profiles can be subsequently decoded locally via the optimal rule of Eq.~\eqref{stddec}. {\bf (b)} Information gain $\Delta PI$ for convolution and divisive normalization in the $L_{{\rm corr}}=0$ and $L_{{\rm corr}} \gg N$ limits (blue and green), respectively.  $\Delta \mathrm{PI}$ achieved by: {\bf (c)} convolution with exponential kernels of varying radii (colorbar; green for the optimal radius), across a range of CI values; {\bf (d)} divisive normalization with a normalization constant obtained via convolution with kernels of varying radii; {\bf (e)} sequential application of convolution (variable radius) and global normalization (sum over all $N$ cells). {\bf (f)} For each correlation length $L_\mathrm{corr}$, the optimal kernel radius differs across the three algorithmic strategies. {\bf (g)} A minimal reaction-diffusion circuit with two diffusible {\it normalizer} species ($N_1$, $N_2$) implements algorithmic operations of (a-f).
    {\bf (h-k)} Same analyses as in {\bf (b-e)}, but using the reaction--diffusion implementation. 
    {\bf (i)} $\Delta \mathrm{PI}$ associated with the steady-state profile $N_1$ while varying its diffusion length $\sqrt{D_1}/N$. 
    {\bf (j)} $\Delta \mathrm{PI}$ associated with $G/N_1$ while varying $\sqrt{D_1}/N$. 
    {\bf (k)} $\Delta \mathrm{PI}$ associated with $N_1/N_1$ while varying $\sqrt{D_1}/N$, with $\sqrt{D_2}/N$ fixed at $\approx 2.23 \gg \sqrt{D_1}/N$. 
    {\bf (l)} For each correlation length, the reaction--diffusion model exhibits optimal diffusion lengths analogous to the optimal kernel radii in (f).
}
    \label{fig:algos_and_mechanism}
\end{figure*}

\textit{RLP in the uncorrelated limit.}
RLP decoding yields largest $\Delta \mathrm{PI}$  when morphogen fluctuations are uncorrelated across cells. As $L_{\rm corr}\rightarrow 0$, optimal \emph{nonlocal} RLP decoding is equivalent to an optimal \emph{local} decoding from an effective readout $g^{\rm eff}_i \propto \sum_{j \in \mathcal{N}_i} g_j$, i.e. a uniform average over a local neighborhood (see SM~\cite{SMnote} Sec.~S3A). This suggests a simple algorithmic strategy: cells first linearly pool information from neighbors to construct a ``smoothed'' profile $g^{\rm eff}$, which can then be locally decoded (Fig.~\ref{fig:algos_and_mechanism}a). 

To investigate the performance of such a pooling strategy, we implement it via spatial convolution, allowing for a systematic variation of the convolutional kernel shape and its spatial extent (``radius''; Fig.~\ref{fig:algos_and_mechanism}a). In the uncorrelated case, we find that $\Delta \mathrm{PI}$ is maximized, though below the optimal bound of Eq.~\eqref{eq:def_deltaPI}, at an intermediate  radius (Fig.~\ref{fig:algos_and_mechanism}b). The RLP decoding yields a robust benefit even beyond the strictly uncorrelated limit: as we vary the correlation length of morphogen fluctuations, substantial $\Delta \mathrm{PI}$ is gained not only at vanishing CI but across a broad range of CI (Fig.~\ref{fig:algos_and_mechanism}c). The precise kernel shape only has mild quantitative effects (in the main text we focus on the exponential kernel; for uniform and Gaussian kernels, see SM Fig.~S2~\cite{SMnote}); what matters most is that the kernel radius is matched to the spatial scale of morphogen fluctuations (Fig.~\ref{fig:algos_and_mechanism}f). 

We also explore more general optimal linear operators $\mathbf{L}$, $\mathbf{g}^{\rm eff}=\mathbf{L}\mathbf{g}$, not restricted to have a convolutional form, which maximize positional information (SM Fig.~S3~\cite{SMnote}). While too complex to inform mechanistic implementations, they provide an existence proof that the optimality bound can be reached with simple linear algorithms, as the theory predicts.


\textit{ALP in the long-correlation limit.}
In contrast to RLP, ALP yields the largest $\Delta \mathrm{PI}$  when morphogen fluctuations are long-range correlated (Fig.~\ref{fig:algos_and_mechanism}c). In the $L_{\rm corr}\rightarrow\infty$ limit -- and assuming that the fluctuation magnitude scales with the mean morphogen levels, $\sigma_{i,E}^2 \propto g_i^2$, as expected from purely extrinsic noise~\cite{swain2002intrinsic} in source-diffusion-degradation models~\cite{grimm2010modelling} -- the ALP decoding also simplifies to a scheme which first transforms the raw morphogen signal into an effective readout $g^{\rm eff}$ that is subsequently locally optimally decoded (see SM~\cite{SMnote} Sec.~S3B). In contrast to the RLP case, where the effective readout was a linear transformation of raw readouts, for the ALP the optimal effective readout is nonlinear,  $g_i^{\text{eff}} \propto {g_i}/G$, with $G={\sum_j g_j}$, i.e., a normalization of each profile by its mean value across positions. Accordingly, the optimal kernel radius for this operation is the system size (Fig.~\ref{fig:algos_and_mechanism}b, second panel). This computation is known as {\it divisive normalization}, a widely-studied operation in sensory neural systems~\cite{carandini2012normalization} where it can be rationalized based on efficient coding theory~\cite{lyu2008nonlinear} (see Sec.~S3 and Fig.~S4 in~\cite{SMnote} for scenario with constant magnitude fluctuations). In our case, divisive normalization emerges as the optimal algorithmic approximation to Bayes-optimal ALP decoding in the fully correlated limit.

Applying divisive normalization to our morphogen ensembles yields an information gain $\Delta \mathrm{PI}$ that qualitatively resembles the behavior of the ALP decoding bound shown in Fig.~\ref{fig:algos_and_mechanism}(d), with performance increasing with correlation strength (high CI).

\textit{Sequential combinations.}
Unlike optimal bounds that are tied to specific assumptions about the structural prior, algorithmic approximations are operations that can be applied sequentially, e.g., convolutional smoothing followed by divisive normalization. The resulting information gain shows a distinct non-monotonic dependence on CI (Fig.~\ref{fig:algos_and_mechanism}e), with the smallest gain occurring in the intermediate correlation length regime that is maximally distinct from the analyzed limiting cases. Nevertheless, the sequential application  of the two algorithms (irrespective of their order) always provides a sizable information gain and outperforms each algorithm applied in isolation; moreover, across a range of CI, the sequential application even provides a synergistic benefit, where the combined benefit is larger than the sum of the individual benefits when the two algorithms are applied in isolation. Such successive transformations of the input signal are reminiscent of multi-layered biological networks that are thought to ``refine'' the positional information~\cite{tkavcik2021many,bruckner2025marrslevelsembryonicdevelopment}. 


\textbf{Level III: Mechanistic implementations.} Our key Level II insight is that the optimal nonlocal decoding can be approximated, across a wide range of morphogen ensembles differing in CI, with the sequential application of two elementary algorithmic operations: spatial convolution and divisive normalization. We next present a simple proof-of-concept reaction-diffusion biochemical network (Fig.~\ref{fig:algos_and_mechanism}g), operating on a static input morphogen profile and transforming it into a refined spatial profile of a new chemical species, $g^{\rm out}(x,t)$, by filtering out much of the short- and long-range fluctuations in the input signal with the help of two intermediary diffusible \emph{normalizer} chemical species $n_1$ and $n_2$~\cite{bruckner_Information_2024}. In this reaction scheme, $g^{\rm out}$ can be seen as a Level III realization of the ``effective'' signal $g^{\rm eff}$ discussed previously, which can be \emph{locally} decoded into precise positional estimates. 

We consider the following reaction-diffusion system:
\begin{eqnarray}
\frac{\partial n_{1}}{\partial t} &=& D_1 \nabla^2 n_1 + \alpha_1 g - \kappa_1 n_1 \\[2mm]
\frac{\partial n_{2}}{\partial t} &=& D_2 \nabla^2 n_2 + \alpha_2 n_1 - \kappa_2 n_2 \\[2mm]
\frac{d g^{\rm out}}{d t} &=& \alpha^{\rm out} n_1 - \kappa^{\rm out} n_2 g^{\rm out}, \label{lasteq}
\end{eqnarray}
with $D_2 \gg D_1$ (see SM~\cite{SMnote} Sec.~S4). $g(x,t)=g(x,0)$ is given as an initial condition; we furthermore assume $n_1(x,0)=n_2(x,0)=0$.

At steady state, the input morphogen profile $g$ is filtered into $n_1$, which naturally acts as its spatially smoothed version; therefore, the first equation describes a mechanism implementing convolution. The fast-diffusing species $n_2$ serves as a global integrator, which can be used to normalize extrinsic fluctuations via an $n_2$-dependent degradation rate of $g^{\rm out}$ in Eq.~\eqref{lasteq}. The final output $g^{\rm out}$ is thus proportional to $n_1/n_2$ at steady state, thereby performing convolution followed by divisive normalization. As shown in Fig.~\ref{fig:algos_and_mechanism}(h-l), this minimal circuit reproduces the $\Delta \mathrm{PI}$ obtained from the algorithmic approximations with a surprising degree of quantitative agreement. Other, more complex reaction schemes should likely be able to either implement identical computations or perhaps instantiate even the non-convolutional transformations of the raw morphogenetic signal discussed in SM Sec.~S3.C~\cite{SMnote}.


\textbf{Conclusions.} A large body of theoretical and experimental work has investigated the limits of reproducibility in cell fate patterns driven by morphogen gradients~\cite{gregor_Probing_2007, rogers_Morphogen_2011, kicheva_Control_2023}. While such pattern formation can be viewed as a cell-autonomous, local, single-step process of morphogen interpretation~\cite{rogers_Morphogen_2011, kicheva_Control_2023}, patterning precision could  be ``refined'' across consecutive steps~\cite{kicheva_Developmental_2015,tkavcik2021many,bruckner2025marrslevelsembryonicdevelopment}, so long as cells are allowed to communicate with each other~\cite{herszterg_Signalingdependent_2025}. 

Here, we formalize the problem of refining positional information through spatial information exchange, which could be mediated by very diverse cell-cell coupling mechanisms. We show that: \emph{(i)} optimal nonlocal decoding can substantially increase positional information in a manner dependent on the correlational information (CI) present in the morphogen ensemble; \emph{(ii)} nonlocal decoding can be approximated by a combination of simple algorithmic transformations across a wide range of morphogen ensembles with different fluctuation correlation structures; \emph{(iii)} such algorithmic transformations can be realized by simple reaction-diffusion-based biochemical networks. 
Taken together, our results demonstrate that the information-theoretic gains of nonlocal decoding are not solely a theoretical curiosity, but could be, in fact, approached by biological systems. Furthermore, our results  provide a framework within which we can attempt to rationalize and interpret the observed multi-layered developmental regulatory cascades~\cite{bruckner2025marrslevelsembryonicdevelopment}.

The optimal outcome of nonlocal decoding depends not only on the statistics of the morphogen ensemble, but also on the assumptions cells can make about their neighbors. This highlights an important dichotomy: with Relative Location Priors (RLP), correlations are detrimental because they make the information pooled by neighbors redundant. In contrast, with Absolute Location Priors (ALP), correlations are beneficial because they reduce the effective number of degrees of freedom in the morphogen ensemble, and thus provide a useful collective signal about the fluctuation that can be utilized to ``undo'' its confounding effect. These qualitative differences in optimal signal processing due to different noise and prior statistics are reminiscent of results in cell signaling~\cite{selimkhanov2014accurate} and in neuroscience~\cite{tkacik_Optimal_2010}. 
In effect, our results show that correlations are not intrinsically good or bad: their utility for reproducible patterning depends on the structural prior about the neighbors that cells might have evolved to rely on in order to interpret nonlocal morphogen signal efficiently.

To date, the literature has focused primarily on the precision of morphogen gradients themselves, paying comparatively little attention to the underlying lattice of cells that ultimately interprets those morphogens. Our study reveals that this cellular lattice--its geometry, its assumptions, and its modes of communication--can be as important as the gradients themselves in shaping positional information. Here, we have captured this effect with idealized priors that assume perfectly reproducible cell positioning and noiseless cell-cell communication.
A natural next step is to relax these assumptions by incorporating finite precision of cell positioning~\cite{deneke_SelfOrganized_2019} and the constraints of noisy communications~\cite{mugler_Limits_2016}, in order to assess how these biological limitations shape the attainable information gains using nonlocal decoding.

\begin{acknowledgments}
This work was supported in part by European Research Council ERC-2023-SyG ``Dynatrans'' grant Nr. 101118866 (GT). We thank Pieter Rein ten Wolde and Vahe Galstyan for stimulating discussions. This work was carried out, in part, at Lucullus, Vienna.
\end{acknowledgments}

\bibliography{references}

@article{bruckner_Information_2024,
  title = {Information Content and Optimization of Self-Organized Developmental Systems},
  author = {Br{\"u}ckner, David B. and Tka{\v c}ik, Ga{\v s}per},
  year = {2024},
  month = jun,
  journal = {Proceedings of the National Academy of Sciences},
  volume = {121},
  number = {23},
  pages = {e2322326121},
  publisher = {Proceedings of the National Academy of Sciences},
  doi = {10.1073/pnas.2322326121}
}

@article{tkacik_Optimal_2010,
  title = {Optimal Population Coding by Noisy Spiking Neurons},
  author = {Tka{\v c}ik, Ga{\v s}per and Prentice, Jason S. and Balasubramanian, Vijay and Schneidman, Elad},
  year = 2010,
  month = aug,
  journal = {Proceedings of the National Academy of Sciences},
  volume = {107},
  number = {32},
  pages = {14419--14424},
  publisher = {Proceedings of the National Academy of Sciences},
  doi = {10.1073/pnas.1004906107}
}

@article{tkavcik2021many,
  title={The many bits of positional information},
  author={Tka{\v{c}}ik, Ga{\v{s}}per and Gregor, Thomas},
  journal={Development},
  volume={148},
  number={2},
  pages={dev176065},
  year={2021},
  publisher={The Company of Biologists Ltd}
}

@article{wolpert1969positional,
  title={Positional information and the spatial pattern of cellular differentiation},
  author={Wolpert, Lewis},
  journal={Journal of theoretical biology},
  volume={25},
  number={1},
  pages={1--47},
  year={1969},
  publisher={Elsevier}
}

@book{cover_Elements_2006,
  title = {Elements of Information Theory},
  author = {Cover, Thomas M. and Thomas, Joy A.},
  year = {2006},
  edition = {2nd ed},
  publisher = {Wiley-Interscience},
  address = {Hoboken, N.J},
  isbn = {978-0-471-24195-9},
  langid = {english},
  lccn = {003.54}
}

@article{deneke_SelfOrganized_2019,
  title = {Self-{{Organized Nuclear Positioning Synchronizes}} the {{Cell Cycle}} in {{Drosophila Embryos}}},
  author = {Deneke, Victoria E. and Puliafito, Alberto and Krueger, Daniel and Narla, Avaneesh V. and Simone, Alessandro De and Primo, Luca and Vergassola, Massimo and Renzis, Stefano De and Talia, Stefano Di},
  year = {2019},
  month = may,
  journal = {Cell},
  volume = {177},
  number = {4},
  pages = {925-941.e17},
  publisher = {Elsevier},
  issn = {0092-8674, 1097-4172},
  doi = {10.1016/j.cell.2019.03.007},
  langid = {english},
  pmid = {30982601}
}

@article{dubuis_Positional_2013,
  title = {Positional Information, in Bits},
  author = {Dubuis, Julien O. and Tka{\v c}ik, Ga{\v s}per and Wieschaus, Eric F. and Gregor, Thomas and Bialek, William},
  year = {2013},
  month = oct,
  journal = {Proceedings of the National Academy of Sciences},
  volume = {110},
  number = {41},
  pages = {16301--16308},
  publisher = {Proceedings of the National Academy of Sciences},
  doi = {10.1073/pnas.1315642110}
}

@misc{erez_NeighborhoodInformed_2025,
  title = {Neighborhood-{{Informed Positional Information}} for {{Precise Cell Identity Specification}}},
  author = {Erez, Michal and Friedman, Roy and Nitzan, Mor},
  year = {2025},
  month = mar,
  primaryclass = {New Results},
  pages = {2025.03.17.643609},
  publisher = {bioRxiv},
  doi = {10.1101/2025.03.17.643609},
  archiveprefix = {bioRxiv},
  chapter = {New Results},
  copyright = {{\copyright} 2025, Posted by Cold Spring Harbor Laboratory. This pre-print is available under a Creative Commons License (Attribution-NonCommercial-NoDerivs 4.0 International), CC BY-NC-ND 4.0, as described at http://creativecommons.org/licenses/by-nc-nd/4.0/},
  langid = {english}
}

@article{gregor_Probing_2007,
  title = {Probing the {{Limits}} to {{Positional Information}}},
  author = {Gregor, Thomas and Tank, David W. and Wieschaus, Eric F. and Bialek, William},
  year = {2007},
  month = jul,
  journal = {Cell},
  volume = {130},
  number = {1},
  pages = {153--164},
  issn = {0092-8674},
  doi = {10.1016/j.cell.2007.05.025}
}

@article{herszterg_Signalingdependent_2025,
  title = {Signaling-Dependent Refinement of Cell Fate Choice during Tissue Remodeling in {{{\emph{Drosophila}}}} Pupal Wings},
  author = {Herszterg, Sophie and Cicolini, Simone and {de Gennes}, Marc and Huang, Anqi and {Matamoro-Vidal}, Alexis and Alexandre, Cyrille and Smith, Matthew and Araujo, Helena and Levayer, Romain and Vincent, Jean-Paul and Salbreux, Guillaume},
  year = {2025},
  month = sep,
  journal = {Developmental Cell},
  issn = {1534-5807},
  doi = {10.1016/j.devcel.2025.08.016}
}

@article{hironaka_Encoding_2012,
  title = {Encoding and Decoding of Positional Information in Morphogen-Dependent Patterning},
  author = {Hironaka, Ken-ichi and Morishita, Yoshihiro},
  year = {2012},
  month = dec,
  journal = {Current Opinion in Genetics \& Development},
  volume = {22},
  number = {6},
  pages = {553--561},
  issn = {1879-0380},
  doi = {10.1016/j.gde.2012.10.002},
  langid = {english},
  pmid = {23200115}
}

@article{kicheva_Control_2023,
  title = {Control of {{Tissue Development}} by {{Morphogens}}},
  author = {Kicheva, Anna and Briscoe, James},
  year = {2023},
  month = oct,
  journal = {Annual Review of Cell and Developmental Biology},
  volume = {39},
  number = {Volume 39, 2023},
  pages = {91--121},
  publisher = {Annual Reviews},
  issn = {1081-0706, 1530-8995},
  doi = {10.1146/annurev-cellbio-020823-011522},
  langid = {english}
}

@article{kicheva_Developmental_2015,
  title = {Developmental {{Pattern Formation}} in {{Phases}}},
  author = {Kicheva, Anna and Briscoe, James},
  year = {2015},
  month = oct,
  journal = {Trends in Cell Biology},
  volume = {25},
  number = {10},
  pages = {579--591},
  issn = {0962-8924},
  doi = {10.1016/j.tcb.2015.07.006}
}

@article{mcgough_Finding_2024,
  title = {Finding the {{Last Bits}} of {{Positional Information}}},
  author = {McGough, Lauren and Casademunt, Helena and Nikoli{\'c}, Milo{\v s} and Aridor, Zoe and Petkova, Mariela D. and Gregor, Thomas and Bialek, William},
  year = {2024},
  month = mar,
  journal = {PRX Life},
  volume = {2},
  number = {1},
  pages = {013016},
  publisher = {American Physical Society},
  doi = {10.1103/PRXLife.2.013016}
}

@article{morishita_Coding_2011,
  title = {Coding {{Design}} of {{Positional Information}} for {{Robust Morphogenesis}}},
  author = {Morishita, Yoshihiro and Iwasa, Yoh},
  year = {2011},
  month = nov,
  journal = {Biophysical Journal},
  volume = {101},
  number = {10},
  pages = {2324--2335},
  issn = {0006-3495},
  doi = {10.1016/j.bpj.2011.09.048}
}

@article{mugler_Limits_2016,
  title = {Limits to the Precision of Gradient Sensing with Spatial Communication and Temporal Integration},
  author = {Mugler, Andrew and Levchenko, Andre and Nemenman, Ilya},
  year = {2016},
  month = feb,
  journal = {Proceedings of the National Academy of Sciences},
  volume = {113},
  number = {6},
  pages = {E689-E695},
  publisher = {Proceedings of the National Academy of Sciences},
  doi = {10.1073/pnas.1509597112}
}

@article{petkova_Optimal_2019,
  title = {Optimal {{Decoding}} of {{Cellular Identities}} in a {{Genetic Network}}},
  author = {Petkova, Mariela D. and Tka{\v c}ik, Ga{\v s}per and Bialek, William and Wieschaus, Eric F. and Gregor, Thomas},
  year = {2019},
  month = feb,
  journal = {Cell},
  volume = {176},
  number = {4},
  pages = {844-855.e15},
  issn = {0092-8674},
  doi = {10.1016/j.cell.2019.01.007}
}

@article{sokolowski2015optimizing,
  title={Optimizing information flow in small genetic networks. IV. Spatial coupling},
  author={Sokolowski, Thomas R and Tka{\v{c}}ik, Ga{\v{s}}per},
  journal={Physical Review E},
  volume={91},
  number={6},
  pages={062710},
  year={2015},
  publisher={APS}
}

@article{hillenbrand2016beyond,
  title={Beyond the French flag model: exploiting spatial and gene regulatory interactions for positional information},
  author={Hillenbrand, Patrick and Gerland, Ulrich and Tka{\v{c}}ik, Ga{\v{s}}per},
  journal={PLoS One},
  volume={11},
  number={9},
  pages={e0163628},
  year={2016},
  publisher={Public Library of Science San Francisco, CA USA}
}

@article{grimm2010modelling,
  title={Modelling the Bicoid gradient},
  author={Grimm, Oliver and Coppey, Mathieu and Wieschaus, Eric},
  journal={Development},
  volume={137},
  number={14},
  pages={2253--2264},
  year={2010},
  publisher={Company of Biologists}
}

@book{mackay_Information_2019,
  title = {Information Theory, Inference, and Learning Algorithms},
  author = {MacKay, David J. C.},
  year = 2019,
  edition = {22nd printing},
  publisher = {Cambridge University Press},
  address = {Cambridge},
  isbn = {978-0-521-64298-9},
  langid = {english}
}

@book{rasmussen_Gaussian_2005,
  title = {Gaussian {{Processes}} for {{Machine Learning}}},
  author = {Rasmussen, Carl Edward and Williams, Christopher K. I.},
  year = 2005,
  month = nov,
  publisher = {The MIT Press},
  doi = {10.7551/mitpress/3206.001.0001},
  copyright = {http://creativecommons.org/licenses/by-nc-nd/4.0/},
  isbn = {978-0-262-25683-4},
  langid = {english}
}

@article{rogers_Morphogen_2011,
  title = {Morphogen {{Gradients}}: {{From Generation}} to {{Interpretation}}},
  shorttitle = {Morphogen {{Gradients}}},
  author = {Rogers, Katherine W. and Schier, Alexander F.},
  year = 2011,
  month = nov,
  journal = {Annual Review of Cell and Developmental Biology},
  volume = {27},
  number = {Volume 27, 2011},
  pages = {377--407},
  publisher = {Annual Reviews},
  issn = {1081-0706, 1530-8995},
  doi = {10.1146/annurev-cellbio-092910-154148},
  langid = {english}
}

@article{swain2002intrinsic,
  title={Intrinsic and extrinsic contributions to stochasticity in gene expression},
  author={Swain, Peter S and Elowitz, Michael B and Siggia, Eric D},
  journal={Proceedings of the National Academy of Sciences},
  volume={99},
  number={20},
  pages={12795--12800},
  year={2002},
  publisher={The National Academy of Sciences}
}

@misc{bruckner2025marrslevelsembryonicdevelopment,
      title={Marr's three levels for embryonic development: information, dynamical systems, gene networks}, 
      author={David B. Brückner and Gašper Tkačik},
      year={2025},
      eprint={2510.24536},
      archivePrefix={arXiv},
      primaryClass={physics.bio-ph},
      url={https://arxiv.org/abs/2510.24536}, 
}

@inproceedings{lyu2008nonlinear,
  title={Nonlinear image representation using divisive normalization},
  author={Lyu, Siwei and Simoncelli, Eero P},
  booktitle={2008 IEEE Conference on Computer Vision and Pattern Recognition},
  pages={1--8},
  year={2008},
  organization={IEEE}
}

@article{carandini2012normalization,
  title={Normalization as a canonical neural computation},
  author={Carandini, Matteo and Heeger, David J},
  journal={Nature reviews neuroscience},
  volume={13},
  number={1},
  pages={51--62},
  year={2012},
  publisher={Nature Publishing Group UK London}
}

@article{deneke2019self,
  title={Self-organized nuclear positioning synchronizes the cell cycle in Drosophila embryos},
  author={Deneke, Victoria E and Puliafito, Alberto and Krueger, Daniel and Narla, Avaneesh V and De Simone, Alessandro and Primo, Luca and Vergassola, Massimo and De Renzis, Stefano and Di Talia, Stefano},
  journal={Cell},
  volume={177},
  number={4},
  pages={925--941},
  year={2019},
  publisher={Elsevier}
}

@article{bauer2021trading,
  title={Trading bits in the readout from a genetic network},
  author={Bauer, Marianne and Petkova, Mariela D and Gregor, Thomas and Wieschaus, Eric F and Bialek, William},
  journal={Proceedings of the National Academy of Sciences},
  volume={118},
  number={46},
  pages={e2109011118},
  year={2021},
  publisher={National Academy of Sciences}
}

@article{selimkhanov2014accurate,
  title={Accurate information transmission through dynamic biochemical signaling networks},
  author={Selimkhanov, Jangir and Taylor, Brooks and Yao, Jason and Pilko, Anna and Albeck, John and Hoffmann, Alexander and Tsimring, Lev and Wollman, Roy},
  journal={Science},
  volume={346},
  number={6215},
  pages={1370--1373},
  year={2014},
  publisher={American Association for the Advancement of Science}
}

@article{piddini2009interpretation,
  title={Interpretation of the wingless gradient requires signaling-induced self-inhibition},
  author={Piddini, Eugenia and Vincent, Jean-Paul},
  journal={Cell},
  volume={136},
  number={2},
  pages={296--307},
  year={2009},
  publisher={Elsevier}
}

@article{erdmann2009role,
  title={Role of spatial averaging in the precision of gene expression patterns},
  author={Erdmann, Thorsten and Howard, Martin and Ten Wolde, Pieter Rein},
  journal={Physical review letters},
  volume={103},
  number={25},
  pages={258101},
  year={2009},
  publisher={APS}
}

@article{cepeda2019estimating,
  title={Estimating information in time-varying signals},
  author={Cepeda-Humerez, Sarah Anhala and Ruess, Jakob and Tka{\v{c}}ik, Ga{\v{s}}per},
  journal={PLoS computational biology},
  volume={15},
  number={9},
  pages={e1007290},
  year={2019},
  publisher={Public Library of Science San Francisco, CA USA}
}

@article{marr1976understanding,
  title={From understanding computation to understanding neural circuitry},
  journal = {AI Memos (MIT)},
  pages = {AIM-357},
  author={Marr, David and Poggio, Tomaso},
  year={1976}
}

@inproceedings{hledik2019tight,
  title={A tight upper bound on mutual information},
  author={Hled{\'\i}k, Michal and Sokolowski, Thomas R and Tka{\v{c}}ik, Ga{\v{s}}per},
  booktitle={2019 IEEE Information Theory Workshop (ITW)},
  pages={1--5},
  year={2019},
  organization={IEEE}
}

@article{tkacik_Positional_2015,
  title = {Positional {{Information}}, {{Positional Error}}, and {{Readout Precision}} in {{Morphogenesis}}: {{A Mathematical Framework}}},
  shorttitle = {Positional {{Information}}, {{Positional Error}}, and {{Readout Precision}} in {{Morphogenesis}}},
  author = {Tka{\v c}ik, Ga{\v s}per and Dubuis, Julien O. and Petkova, Mariela D. and Gregor, Thomas},
  year = {2015},
  month = jan,
  journal = {Genetics},
  volume = {199},
  number = {1},
  pages = {39--59},
  issn = {0016-6731},
  doi = {10.1534/genetics.114.171850},
  pmcid = {PMC4286692},
  pmid = {25361898}
}

@article{zagorski_Decoding_2017,
  title = {Decoding of Position in the Developing Neural Tube from Antiparallel Morphogen Gradients},
  author = {Zagorski, Marcin and Tabata, Yoji and Brandenberg, Nathalie and Lutolf, Matthias P. and Tka{\v c}ik, Ga{\v s}per and Bollenbach, Tobias and Briscoe, James and Kicheva, Anna},
  year = {2017},
  month = jun,
  journal = {Science},
  volume = {356},
  number = {6345},
  pages = {1379--1383},
  publisher = {American Association for the Advancement of Science},
  doi = {10.1126/science.aam5887}
}

@MISC{SMnote,
NOTE = {See Supplemental Material at http://XXX for details on the mathematical derivation of the variational algorithm and for supplementary figures and table}
}

\clearpage

\widetext
\clearpage
\begin{center}
\textbf{%
  {\LARGE Supplemental Material}\\ \vspace*{1.5mm}
  {Nonlocal decoding of positional and correlational information during development}
} \\
\vspace*{5mm}
Alex Chen Yi Zhang, Pablo Mateu Hoyos, David Br\"uckner, Ga\v{s}per Tka\v{c}ik
\vspace*{10mm}
\end{center}

\setcounter{equation}{0}
\setcounter{figure}{0}
\setcounter{table}{0}
\setcounter{page}{1}
\setcounter{section}{0}
\setcounter{page}{1}
\makeatletter
\renewcommand{\theequation}{S\arabic{equation}}
\renewcommand{\thefigure}{S\arabic{figure}}
\renewcommand{\thetable}{S\arabic{table}}
\renewcommand{\thesection}{S\arabic{section}}
\renewcommand{\thepage}{S\arabic{page}}

\onecolumngrid

This Supplemental Material provides additional theoretical and methodological details supporting the main text. We begin with a brief recap of positional information (PI), correlational information (CI), and the framework of optimal local decoding, followed by the introduction of optimal nonlocal decoding (Section~\ref{sec::intro_and_definitions},\ref{sec:CI},\ref{sec:nonlocal}). We then present detailed derivations of the Relative Locations Prior (RLP) and Absolute Locations Prior (ALP) bounds on the positional information gain enabled by nonlocal decoding (Sections~\ref{sec:RLP},\ref{sec:ALP}). Next, we develop and analyze algorithmic approximations to these optimal decoding strategies (Section~\ref{sec:algorithmmic_limits}). Finally, we describe a minimal reaction-diffusion network that implements these computations, together with steady-state solutions and simulation details (Section~\ref{sec:reaction_diffusion}).

\vspace{1em}

\noindent\textbf{Contents}

\vspace{0.5em}

\begin{list}{}{\leftmargin=1.5em \itemsep=0.25em}

\item \textbf{S1. A brief recap of PI and local decoding}

\item \textbf{S2. CI and nonlocal decoding}
  \begin{list}{}{\leftmargin=1.5em \itemsep=0.15em}
    \item A. Correlational Information
    \item B. Optimal Nonlocal Decoding
    \item C. Relative Locations Prior (RLP)
    \item D. Absolute Locations Prior (ALP)
  \end{list}

\item \textbf{S3. Algorithmic approximations to optimal nonlocal decoding}
  \begin{list}{}{\leftmargin=1.5em \itemsep=0.15em}
    \item A. RLP for uncorrelated noise
    \item B. ALP for perfectly correlated noise
    \item C. More complex algorithmic approximations
    \item Approximations for Relative Locations Prior (RLP)
    \item Approximations for Absolute Locations Prior (ALP)
  \end{list}

\item \textbf{S4. Reaction--diffusion model: steady state solution and simulation details}
  \begin{list}{}{\leftmargin=1.5em \itemsep=0.15em}
    \item A. Steady-state solution and interpretation
    \item B. Simulation details
  \end{list}

\item \textbf{Supplemental Figures}

\item \textbf{References}

\end{list}

\vspace{1em}

\clearpage

\section{A brief recap of PI and {\it local} decoding}\label{sec::intro_and_definitions}
To introduce the theoretical framework of positional information (PI), consider a morphogen profile $g(x)$ along a one-dimensional embryo. $g(x)$ is subject to stochastic fluctuations arising from both intrinsic molecular noise and embryo-to-embryo variability~\cite{gregor_Probing_2007}. At each position $x$, this defines a conditional distribution $P(g|x)$, often well approximated by a Gaussian distribution with mean profile $\bar{\mathbf{g}}(x)$ and variance  $\sigma_g^2(x)$~\cite{dubuis_Positional_2013}.

PI quantifies how much information the morphogen level conveys about position. Formally, it is defined as the mutual information~\cite{cover_Elements_2006} between the position $X$ and morphogen concentration $G$:
\begin{equation}
\label{eq:PI_def_1}
    \mathrm{PI}:= I(X,G) = \int dx P(x) \int dg P(g|x) \log_2 \frac{P(g|x)}{P(g)}
\end{equation}
If $\mathrm{PI} = \log_2 N \,\,\mathrm{bits}$ for an embryo consisting of $N$ cells, then the morphogen gradient contains enough information to uniquely specify the identity of every cell.

Beyond this interpretation as a measure of statistical dependency, PI admits a natural decoding interpretation.
In this view, cells infer their positions from local morphogen readouts by performing a form of statistical inference~\cite{morishita_Coding_2011, hironaka_Encoding_2012}. If this process is close to Bayes-optimal, the estimated position is given by the maximum {\it a posteriori} (MAP) estimator $\hat{x} = {\rm argmax}_{x^{\ast}} P(x^{\ast}|g)$ that maximizes the posterior over the implied position $x^{\ast}$. The precision of this inference is quantified by the mutual information $I(X,\hat{X})$. If the posteriors are unimodal and the small noise approximation holds~\cite{petkova_Optimal_2019, tkacik_Positional_2015}, positional information can be expressed as:
\begin{equation}
    PI\simeq I(X,\hat{X}) \simeq - \Big\langle \log_2 \frac{ \sqrt{2\pi e \sigma_x^2(x)}}{L} \Big\rangle_x 
\label{eq:SM_PI_and_poserrs}
\end{equation}
where $L$ is the embryo length and the {\it positional error}, $\sigma_x^2(x) = \langle(x-x^{\ast})^2\rangle_{P(x^{\ast}|x)}$, is the mean squared error of the inferred position $\hat{x}$ given morphogen readouts at the position $x$. For a uniform prior over positions, the positional error coincides with the variance of the maximum-likelihood estimator based on the local likelihood $P(g|x^{\ast})$.

The precision of any unbiased estimator is bounded from below by the Cramer-Rao bound~\cite{cover_Elements_2006}:
\begin{equation}
    \sigma_x^2(x) \geq \mathcal{I}^{-1}(x)
    \label{eq:cramer_rao}
\end{equation}
where $\mathcal{I}(x)$ is the Fisher information. For Gaussian morphogen fluctuations, it can be approximated as ~\cite{tkacik_Positional_2015, petkova_Optimal_2019}:
\begin{equation}
    \mathcal{I}(x) =  - \mathbb{E}_{P(g|x)}\left[ \frac{\partial^2 \log P(g|x)}{\partial x^2} \right] \simeq \frac{(\bar{g}'(x))^2}{\sigma_g^2(x)}
    \label{eq:fisher}
\end{equation}
where $\bar{g}'(x)$ is the spatial derivative of the mean profile, and $\sigma_g^2(x)$ is the variance of morphogen level fluctutations at location $x$.

\section{CI and nonlocal decoding}\label{sec:CI_and_nl_decoding}
\subsection{Correlational Information}\label{sec:CI}
The framework described above assumes that cells make independent inferences about their position by relying only on local morphogen readouts. This is equivalent to neglecting the spatial correlations in the morphogen signal fluctuations (which could be strong in the case of dominant extrinsic noise).
To capture correlated variability, we model the ensemble of morphogen profiles $\mathbf{g}=\{g_1,\ldots,g_N\}$ over $N$ cells at locations $\mathbf{x}=\{x_1, \ldots, x_N\}$ as a multivariate Gaussian $P(\mathbf{g}|\mathbf{x}) = \mathbf{g} \sim \mathcal{N}(\bar{\mathbf{g}},\mathbf{C})$,

\begin{equation}
    P(\mathbf{g}|\mathbf{x}) = \frac{1}{\sqrt{(2\pi)^N |\mathbf{C}|}} \exp \{- \frac{\chi}{2} \}
\end{equation}
with
\begin{equation}
    \chi = \sum_{i,j = 1}^N (g_i - \bar{g}_i)[C^{-1}]_{ij} (g_j - \bar{g}_j)
\end{equation}

where $\bar{g}_i \equiv \bar{g}(x_i)$ is the mean morphogen profile, and the covariance matrix $\mathbf{C}$ encodes spatial correlations. Throughout this work, we consider a simple exponential gradient $\bar{g}(x_i) = \mu\exp(-x_i/\lambda)$ with length scale $\lambda$ and magnitude $\mu$, and we parametrize $\mathbf{C}$ in the following way:
\begin{equation}
    C_{ij} = v(x_i)\left[\exp\Big(-\frac{|x_i - x_j|}{L_{\rm corr}}\Big)\sigma_E^2 + \delta_{ij}\sigma_I^2\right]v(x_j) = \mathcal{C}(x_i,x_j)
    \label{eq:covariance_matrix}
\end{equation}
which is defined by the covaraince kernel~\cite{rasmussen_Gaussian_2005} $\mathcal{C}(\cdot,\cdot)$. The parameter $L_{\rm corr}$ sets the spatial range of the correlations, while $\sigma_E/\sigma_I$ controls their relative strength. The function $v(\cdot)$ modulates the local amplitude of fluctuations, such that the variance of morphogen levels at position $x_i$ is $\sigma_g^2(x_i) = v^2(x_i)[\sigma_E^2 + \sigma_I^2]$. In the main text, we focus on the choice $v(x_i) \propto \bar{g}(x_i)$, as one would expect in the case of purely extrinsic noise~\cite{swain2002intrinsic} in source-diffusion-degradation models~\cite{grimm2010modelling}. For comparison, we also consider the alternative case $v(x_i) = 1$.

The amount of information carried by the spatial correlations can be quantified using {\it Correlational Information} (CI)~\cite{bruckner_Information_2024}. CI is defined as the Kullback-Leibler divergence~\cite{cover_Elements_2006} between the full joint distribution of the morphogen levels across the embryo and the product of the corresponding marginal distributions:
\begin{equation}
    \mathrm{CI} := \frac{1}{N}D_{\text{KL}}[P(\mathbf{g}|\mathbf{x})||\prod_{i=1}^N P(g_i|x_i)]
    \label{eq:CI}
\end{equation}
By construction, CI vanishes when morphogen fluctuations are independent across positions, and increases as correlations become stronger or longer-ranged. This behavior is illustrated in Figure~\ref{fig:information_param_scan}(a), where CI grows monotonically with both $L_{\rm corr}$ and $\sigma_E/\sigma_I$.

\subsection{Optimal Nonlocal Decoding}\label{sec:nonlocal}
We now generalize the local decoding framework to allow cells to exploit nonlocal information. Specifically, we consider the problem in which each cell $i$ seeks to infer its own position $x_i$ not only from local measurement $g_i$, but also from additional measurements $\{g_j\}_{j \in \mathcal{N}_i}$ obtained from other cells in the neighborhood $\mathcal{N}_i$. This neighborhood may be limited in spatial extent or, in the extreme case, include the entire embryo. 
For notational simplicity, we denote by $\mathbf{g}$ the full set of morphogen measurements accessible to a cell $i$, i.e. $\{g_j\}_{j \in \mathcal{N}_i}$.

The inferred position is taken to be the maximum {\it a posteriori} estimate:
\begin{equation}
    \hat{x}_i = {\rm argmax}_{x^{\ast}_i} P(x^{\ast}_i|\mathbf{g})
\end{equation}
The posterior distribution can be expressed as
\begin{align}
P(x^{\ast}_i|\mathbf{g}) &= \int d\mathbf{x}^{\ast}_{\neg i} P(\mathbf{x}^{\ast}|\mathbf{g}) =  \quad \text{Bayes th.} \nonumber \\
&= \int d\mathbf{x}^{\ast}_{\neg i} \frac{P(\mathbf{g}|\mathbf{x}^{\ast})P(\mathbf{x})}{P(\mathbf{g})} \propto \nonumber \\
& \propto P(x^{\ast}_i) \int d\mathbf{x}^{\ast}_{\neg i} P(\mathbf{g}|\mathbf{x}^{\ast})P(\mathbf{x}^{\ast}_{\neg i}|x^{\ast}_i)
\label{eq:gen_posterior}
\end{align}
where $\mathbf{x}^{\ast}_{\neg i}$ denotes the vector of implied positions of all cells other than $i$, and $P(\mathbf{g}|\mathbf{x}^{\ast})$ is the joint likelihood of morphogen readouts given the full (assumed) cell-position vector, $\mathbf{x}^{\ast}$.
Thus, two key factors determine the limits to non-local positional decoding: \emph{(i)}, the correlation structure of morphogen fluctuations as captured by $P(\mathbf{g}|\mathbf{x}^{\ast})$; and \emph{(ii)}, the ``structural prior,'' $P(\mathbf{x}^{\ast}_{\neg i}|x^{\ast}_i)$, which formalizes what a cell $i$ can assume about the positions of other ($j\neq i$) cells that contribute their non-local morphogen readouts.

Typically, the integral over the structural prior in Eq.~(\ref{eq:gen_posterior}) will be analytically intractable.  To make progress and simplify the calculations, we analyze two special cases that represent limiting forms of nonlocal decoding, as we detail below. In both cases, our goal is to quantify the improvement in positional inference enabled by nonlocal information. As in the local decoding framework, we characterize inference precision through the mutual information between the true position and inferred positions $I(X,\hat{X}_{\text{nl}})$, where ``$\text{nl}$'' stands for decoding with access to nonlocal information; and compare it to the local decoding baseline $I(X,\hat{X}_{\text{l}})$. 
We define
\begin{equation}
    \Delta \mathrm{PI} := I(X,\hat{X}_{\text{nl}}) - I(X,\hat{X}_{\text{l}})
\label{eq:SM_def_deltaPI}
\end{equation}
as the gain in positional information, in bits, due to the optimal utilization of nonlocal information in the form of morphogen readouts at additional positions $j \in \mathcal{N}_i$, which we assume can be communicated in an error-free fashion to position $i$. In reality, any such communication would incur extra transmission noise, limiting its utility~\cite{mugler_Limits_2016}, a constraint we neglect here to compute optimal bounds, yet essential for a more complete future theory. With this caveat in mind, we proceed below to show that $\Delta \mathrm{PI}$ strongly depends on the form of the structural prior, by focusing on two ideal limits: the Relative (RLP) and Absolute Locations Prior (ALP), respectively.

\subsection{Relative Locations Prior (RLP)}\label{sec:RLP}
We first consider {\it RLP}, which represents the limit in which each cell has precise knowledge of the relative positions of all other cells contributing nonlocal morphogen information. Formally, this structural prior can be expressed as: 
\begin{equation}
    P_{\rm RLP}(\mathbf{x}^{\ast}_{\neg i}|x^{\ast}_i) = \prod_{j} \delta (x^{\ast}_j - (x^{\ast}_i+d_{ij}))
\end{equation}
where $d_{ij} = x_j - x_i$.
Substituting this into the general expression for the posterior Equation~\ref{eq:gen_posterior} yields:
\begin{equation}
    P(x^{\ast}_i|\mathbf{g}) \propto P(x^{\ast}_i) P(\mathbf{g}|\ldots (x^{\ast}_i + d_{i,i-1}), x^{\ast}_i,(x^{\ast}_i + d_{i,i+1})\ldots)
\end{equation}
If $P(x^{\ast}_i)$ is uniform, the MAP estimate of $x_i$ reduces to maximizing the log-likelihood.
\begin{align}
    \mathcal{L}(x^{\ast}_i) & = \log  P(\mathbf{g}|\ldots (x^{\ast}_i + d_{i,i-1}), x^{\ast}_i,(x^{\ast}_i + d_{i,i+1})\ldots) \nonumber \\
    & = -\frac{1}{2}\log|C(x_i^{\ast})| - \frac{1}{2} \sum_{j,k} \bigl( g_j - \bar{g}(x^{\ast}_i + d_{ij})\bigr) [C(x_i^{\ast})^{-1}]_{jk} \bigl( g_k - \bar{g}(x^{\ast}_i + d_{ik})\bigr)
\label{eq:RLP_likelihood}
\end{align}
 
For notational simplicity, let us write $P(\mathbf{g}|\ldots (x^{\ast}_i + d_{i,i-1}), x^{\ast}_i,(x^{\ast}_i + d_{i,i+1})\ldots) = P(\mathbf{g}|x^{\ast}_i)$.
To quantify the precision of this inference, we compute the Fisher information associated with maximizing $P(\mathbf{g}|x^{\ast}_i)$:
\begin{align}
    \mathcal{I}_{\rm RLP}(x_i)  = & - \mathbb{E}_{P(\mathbf{g}|x_i)}\left[ \frac{\partial^2 \log P(\mathbf{g}|x_i)}{\partial x_i^2} \right] \stackrel{Tr. inv.}{=} \nonumber \\
     = & \mathbb{E}_{P(\mathbf{g}|x_i)}\left[ \frac{1}{2} \sum_{j,k} \frac{\partial^2}{\partial x_i^2}  \bigl( g_j - \bar{g}(x_i + d_{ij})\bigr) [C^{-1}]_{jk} \bigl( g_k - \bar{g}(x_i + d_{ik})\bigr) \right] = \nonumber \\
     = & \mathbb{E}_{P(\mathbf{g}|x_i)}\left[ \frac{2}{2} \sum_{j,k } - \frac{\partial}{\partial x_i} \bar{g}'(x_i + d_{ij}) [C^{-1}]_{jk} \bigl( g_k - \bar{g}(x_i + d_{ik})\bigr) \right] = \nonumber \\
     = & \mathbb{E}_{P(\mathbf{g}|x_i)}\left[ -\sum_{j} \bar{g}''(x_i + d_{ij}) \sum_{k \in \mathcal{N}_i} [C^{-1}]_{jk} \bigl( g_k - \bar{g}(x_i + d_{ik})\bigr) \right] \nonumber\\
     & \qquad\qquad +\mathbb{E}_{P(\mathbf{g}|x_i)}\left[ \sum_{j} \bar{g}'(x_i + d_{ij}) \sum_{k } [C^{-1}]_{jk} \bar{g}'(x_i + d_{ik}) \right]  = \nonumber \\
     = & \sum_{j,k} \bar{g}'(x_i + d_{ij}) [C^{-1}]_{jk} \bar{g}'(x_i + d_{ik})
\end{align}
In the first step, we neglect the spatial variability of the variance of morphogen fluctuations; in other words, we take the covariance kernel from Eq.~\ref{eq:covariance_matrix} to be translationally invariant. In the second-to-last step, the term containing $\bigl( g_k - \bar{g}(x_i + d_{ik})\bigr)$ becomes zero after averaging. 
In the case each cell $i$ has access to information from a limited neighborhood $\mathcal{N}_i$, we can then write the positional errors associated with RLP nonlocal decoding as:
\begin{equation}
    \sigma_{x,\text{RLP}}^2(x_i) \simeq \mathcal{I}_{\rm RLP}^{-1}(x_i) \simeq \left ( \sum_{j,k \in \mathcal{N}_i} \bar{g}'(x_j) [C_{\mathcal{N}_i}^{-1}]_{jk} \bar{g}'(x_k) \right)^{-1}
    \label{eq:SM_rlp_poserrs}
\end{equation}
Here, $[C_{\mathcal{N}_i}^{-1}]_{jk}$ is the $(j,k)$ element of the inverse covariance matrix of morphogen fluctuations restricted to the neighborhood $\mathcal{N}_i$. Figure~\ref{fig:information_param_scan}(c) shows $\Delta PI_{\rm RLP}$ computed with Equation~\ref{eq:SM_rlp_poserrs}as a function of $L_{\rm corr}$ and $\sigma_E/\sigma_I$ values.

\subsection{Absolute Locations Prior (ALP)}\label{sec:ALP}
We now consider the opposite limiting case, the {\it Absolute Locations Prior (ALP)}. In this scenario, cell $i$ is assumed to know the absolute positions of all other cells contributing nonlocal morphogen measurements, while its own position remains unknown. Formally, the structural prior under ALP is given by:
\begin{equation}
    P_{\text{ALP}}(\mathbf{x}^{\ast}_{\neg i}|x^{\ast}_i) = P(\mathbf{x}^{\ast}_{\neg i})= \prod_{j \in \mathcal{N}_i} \delta (x^{\ast}_j - x_j).
\end{equation}
which fixes the implied positions of all cells other than $i$ to their true locations.
Substituting this prior in Equation~\ref{eq:gen_posterior} yields:
\begin{equation}
    P(x^{\ast}_i|\mathbf{g}) = \frac{P(x^{\ast}_i)}{P(\mathbf{g})}P(g_i|\mathbf{g}_{\neg i}, \mathbf{x}_{\neg i}, x^{\ast}_i)P(\mathbf{g}_{\neg i}|\mathbf{x}_{\neg i})
\end{equation}
If $P(x^{\ast}_i)$ is uniform, the inference problem reduces to maximizing the conditional likelihood $P(g_i|\mathbf{g}_{\neg i}, \mathbf{x}_{\neg i}, x^{\ast}_i)$. Because the joint distribution of morphogen levels is multivariate Gaussian, this conditional likelihood is itself a univariate Gaussian distribution. Specifically, $P(g_i|\mathbf{g}_{\neg i}, \mathbf{x}_{\neg i}, x^{\ast}_i) = \mathcal{N}(g_i|\tilde{g}(x^{\ast}_i), \tilde{\sigma}(x^{\ast}_i)^2)$ where:

\begin{align}
\tilde{g}(x^{\ast}_i) &= \bar{g}(x^{\ast}_i) + C_{i, \neg i}[\mathbf{C}_{\neg i, \neg i}]^{-1}(\mathbf{g}_{\neg i} - \bar{\mathbf{g}}(\mathbf{x}_{\neg i})) \\
\tilde{\sigma}^2(x^{\ast}_i) &= C_{i,i} - C_{i, \neg i}[\mathbf{C}_{\neg i, \neg i}]^{-1}C_{\neg i, i}
\label{eq:conditional_gaussian_parameters}
\end{align}
The expression for $\tilde{\sigma}^2(x^{\ast}_i)$ is the Schur complement of $C_{ii}$ in the matrix $\mathbf{C}$. By using a useful property of the Schur complement, we can write:
\begin{equation}
    \tilde{\sigma}^2(x^{\ast}_i) = ([C^{-1}]_{ii})^{-1}
\end{equation}

\begin{figure}
    \centering
    \includegraphics[width=0.7\linewidth]{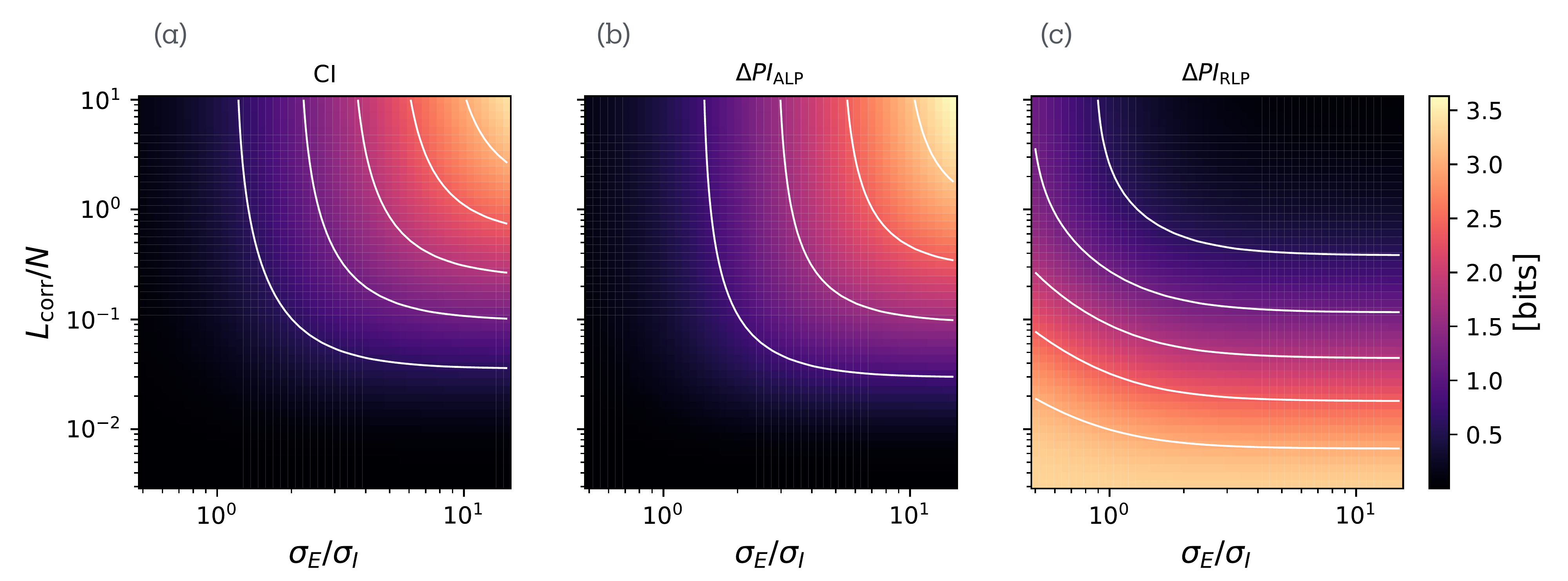}    
    \caption{$CI$, $\Delta PI_{{\rm RLP}}$, and $\Delta PI_{{\rm ALP}}$ as a function of the correlation length of the fluctuations and the ratio between the magnitude of the ``{\it extrinsic}'' and ``{\it intrinsic}'' parts of the fluctuations. The results shown here are for the case in which the neighborhood corresponds to the full embryo size i.e., $|\mathcal{N}_i|=N$.}
    \label{fig:information_param_scan}
\end{figure}

Similarly to what we did above, we compute the Fisher information for the maximization of this Gaussian likelihood:
\begin{equation}
\label{eq:fisher_ALP}
    \mathcal{I}_{\rm ALP}(x_i) = - \mathbb{E}_{P(g_i|\mathbf{g}_{\neg i}, \mathbf{x}_{\neg i}, x_i)}\left[ \frac{\partial^2 \log P(g_i|\mathbf{g}_{\neg i}, \mathbf{x}_{\neg i}, x_i)}{\partial x_i^2}  \right] \simeq \frac{(\tilde{g}'(x_i))^2}{\tilde{\sigma}^2(x_i)} 
\end{equation}
This leads to the following positional errors for ALP:
\begin{equation}
    \sigma_{x,{\rm ALP}}^2(x_i)\simeq \mathcal{I}_{\rm ALP}^{-1}(x_i) \simeq \left( \bar{g}'(x_i)^2 [C_{\mathcal{N}_i}^{-1}]_{ii}\right)^{-1}
\end{equation}
Combining this result with Equation~\ref{eq:SM_def_deltaPI}, we can write a compact expression for the positional information gain $\Delta PI_{\rm ALP}$ granted by ALP nonlocal decoding, when each cell $i$ has access to information from a neighborhood $\mathcal{N}_i$:
\begin{equation}
\label{eq:delta_PI_ALP}
    \Delta PI_{\rm ALP} = \frac{1}{2N}\sum_{i=1}^N \left[\log_2 [C_{\mathcal{N}_i}]_{ii} + \log_2 [C_{\mathcal{N}_i}^{-1}]_{ii}  \right]
\end{equation}
Figure~\ref{fig:information_param_scan}(b) shows $\Delta PI_{\rm ALP}$ as a function of the correlation length $L_{\rm corr}$ and the extrinsic-to-extrinsic noise ratio $\sigma_E/\sigma_I$ values.

\section{Algorithmic approximations to optimal nonlocal decoding}\label{sec:algorithmmic_limits}

\subsection{RLP for uncorrelated noise}\label{sec:RLP_algo_limit}
Stemming from the result that RLP yields the largest $\Delta PI$ when morphogen fluctuations are uncorrelated across cells, we derive in the limit $L_{\rm corr} \rightarrow 0$, the form of the decoded position $\hat{x}_i$, to find a simple algorithmic approximation to the optimal decoding. In particular, we consider a simplified case in which $C_{jk} = \delta_{jk} \sigma^2$. The decoded position is then inferred by solving the following optimization problem:
\begin{equation}
     \hat{x}_i = {\rm argmin}_{x^{\ast}_i} [-\mathcal{L}(x^{\ast}_i)] = {\rm argmin}_{x^{\ast}_i} \sum_{j,k \in \mathcal{N}_i} \bigl( g_j - \bar{g}(x^{\ast}_i + d_{ij})\bigr) \bigl( g_k - \bar{g}(x^{\ast}_i + d_{ik})\bigr) \delta_{jk} \sigma^2
\end{equation}
Then:
\begin{equation}
    -\mathcal{L}(x^{\ast}_i) \propto \sigma ^2\sum_{j \in \mathcal{N}_i} \bigl[ g_j - \bar{g}(x^{\ast}_i+d_{ij}) \bigr]^2
    \label{eq:regression_loss}
\end{equation}
To make further analytical progress, we assume that the $\mathcal{N}_i$ is a small and symmetric neighborhood around cell $i$. In this case we can linearize the profile around $x_i^{\ast}$: $\bar{g}(x^{\ast}_i + d_{ij}) \simeq \bar{g}(x^{\ast}_i) + d_{ij} \,\bar{g}'(x^{\ast}_i)$. Defining $a \equiv \bar{g}(x_i)$ and $b \equiv \bar{g}'(x_i)$, the above quantity to be maximized becomes:
\begin{equation}
    -\mathcal{L}(x^{\ast}_i) \propto \sum_{j \in \mathcal{N}_i} \bigl[ g_j - (a+bd_{ij}) \bigr]^2
\end{equation}
This expression is identical to the objective function of an ordinary least squares problem, corresponding to the linear regression~\cite{mackay_Information_2019}:
\begin{equation}
    g_j = a + b\,d_{ij} + \epsilon_{j}
\end{equation}
where $\epsilon_{j}$ denotes the readout errors.
In matrix form, this can be written as $\mathbf{g} = D\mathbf{\theta} + \mathbf{\epsilon}$. Here, $\theta = (a,b)^T=(\bar{g}(x^{\ast}_i),\bar{g}'(x^{\ast}_i))^T$ and $D$ is a $|\mathcal{N}_i|\times2$ matrix where the first column is made of ones and the second column contains the $d_{ij}$'s.

The ordinary least squares solutions is given by $\mathbf{\hat{\theta}} = (D^T D)^{-1} D^T\mathbf{g}$. If the neighborhood $\mathcal{N}_i$ is symmetric around cell $i$, then $\sum_{j \in \mathcal{N}_i} d_{ij} = 0$. And the estimate of $\theta_1$ simplifies to:
\begin{equation}
    \hat{\theta}_1 = \bar{g}(\hat{x}_i) = \frac{1}{|\mathcal{N}_i|} \sum_{j \in \mathcal{N}_i} g_j \equiv g_i^{\rm eff}
\end{equation}

Therefore, in the limit we have considered, RLP nonlocal decoding reduces to choosing the value of $x^{\ast}_i$ such that $\bar{g}(x^{\ast}_i) = g_i^{\rm eff}$. This suggests a simple algorithmic strategy: cells first linearly pool information from neighbors to construct a ``smoothed'' profile $\mathbf{g}^{\rm eff}$, and then perform standard local decoding on this smoothed signal.

Because of the approximations made during the derivation, this uniform averaging need not be optimal. Motivated by this observation, we also consider weighted pooling schemes in which morphogen measurements are combined using spatial kernels, such as exponential or Gaussian kernels. These convolutional filters implement the same qualitative strategy, while allowing different weights to be assigned to neighbors at different distances. Figure~\ref{fig:different_kernels} (left column), for a comparison between the different kernels.

\subsection{ALP for perfectly correlated noise}\label{sec:ALP_algo_limit}

We have shown that ALP decoding works best at high correlation lengths. Here, we take the limit in which the correlation length diverges $L_{\rm corr}\rightarrow \infty$, and ask whether this limit suggests a simple algorithmic strategy that approaches the ALP optimal bound.

We recall the general form of the covariance matrix introduced in Equation~\ref{eq:covariance_matrix}. Focusing on correlated fluctuations, we neglect the small diagonal contribution proportional to $\sigma_I$ and write
\begin{equation}
    C_{ij} = v(x_i)\left[\exp\Big(-\frac{|x_i - x_j|}{L_{\rm corr}}\Big)\sigma_E^2 \right]v(x_j) \equiv v_i K_{ij}v_j
\end{equation}
where $K_{ij}$ captures the spatial correlations, controlled by the parameter $L_{corr}$, while the function $v(x)$ encodes the spatial modulation of the local variance.

The terms appearing in Eqs.~\ref{eq:conditional_gaussian_parameters} can be expressed as:
\begin{align}
    [C_{\neg i, \neg i}^{-1}]_{jk} & =  \frac{[K_{\neg i, \neg i}^{-1}]_{jk}}{v_j v_k} \\
    [C_{i,\neg i}]_j & =  v_i v_j K_{ij} \xrightarrow[L_{corr} \to \infty]{} v_iv_j
\end{align}
Plugging these expressions into the original equations yields:
\begin{align}
   \tilde{\sigma}^2(x^{\ast}_i) &= C_{i,i} - C_{i, \neg i}[C_{\neg i, \neg i}]^{-1}C_{\neg i, i} \nonumber \\
   & = v_i^2-v_i^2\sum_{j,k \neq i}v_j \frac{[K_{\neg i, \neg i}^{-1}]_{jk}}{v_j v_k} v_k \nonumber \\
   & = v_i^2-v_i^2\sum_{j,k \neq i} [K_{\neg i, \neg i}^{-1}]_{jk} = 0
\end{align}
here, we have used that $\sum_{j,k \neq i} [K_{\neg i, \neg i}^{-1}]_{jk} = 1, \,\, \forall i$, for an exponential covariance kernel~\cite{rasmussen_Gaussian_2005}. Therefore, the variance of the conditional distribution is zero in the limit of infinite correlation length. This result is intuitive: when fluctuations are perfectly correlated across space, there is effectively a single fluctuating degree of freedom. As a consequence, maximizing the conditional distribution reduces to finding the position $\hat{x}_i$, such that $g_i = \tilde{g}(\hat{x}_i)$. The conditional mean from Eqs.~\ref{eq:conditional_gaussian_parameters} can be written as:

\begin{equation}
    \tilde{g}(\hat{x}_i) = \bar{g}(\hat{x}_i) + v_i \sum_{j,k \neq i}v_j \frac{[K_{\neg i, \neg i}^{-1}]_{jk}}{v_j v_k} (g_k - \bar{g}(x_k))
\end{equation}
In the limit $L_{\rm corr}\rightarrow \infty$, morphogen levels at all locations within a single embryo become perfectly correlated and can be written as:
\begin{equation}
    \mathbf{g} = a\mathbf{v} + \bar{g}(\mathbf{x}) \qquad \text{with} \,\, a\sim\mathcal{N}(0,1)
\end{equation}
and in the case where $v(x) =  \gamma \bar{g}(x)$ ($\gamma$ here is a simple proportionality constant), we have the following identity:
\begin{equation}
    \frac{\sum_j g_j}{\sum_j\bar{g}(x_j)} = (1+\gamma \sigma)
\end{equation}
Using these identities, we can write:
\begin{align}
    g_i = \tilde{g}(\hat{x}_i) & = \bar{g}(\hat{x}_i) + v_i \sum_{j,k \neq i}v_j \frac{[K_{\neg i, \neg i}^{-1}]_{jk}}{v_j v_k} a v_k  = \nonumber \\
    & = \bar{g}(\hat{x}_i) + av_i \sum_{j,k \neq i} [K_{\neg i, \neg i}^{-1}]_{jk} = \nonumber \\
    & = \bar{g}(\hat{x}_i) \Big( 1 +a\frac{v_i}{\bar{g}(\hat{x}_i)} \Big) = \bar{g}(\hat{x}_i) (1 + a\sigma) = \nonumber \\
    & = \bar{g}(\hat{x}_i)\frac{\sum_j g_j}{\sum_j\bar{g}(x_j)}
\end{align}
From the final equality, we conclude that the decoded position must satisfy:
\begin{equation}
    \bar{g}(\hat{x}_i) =  \frac{g_i}{\sum_j g_j} \sum_j \bar{g}(x_j) \equiv g^{\rm eff}_i
\end{equation}

\begin{figure}
    \centering
    \includegraphics[width=0.8\linewidth]{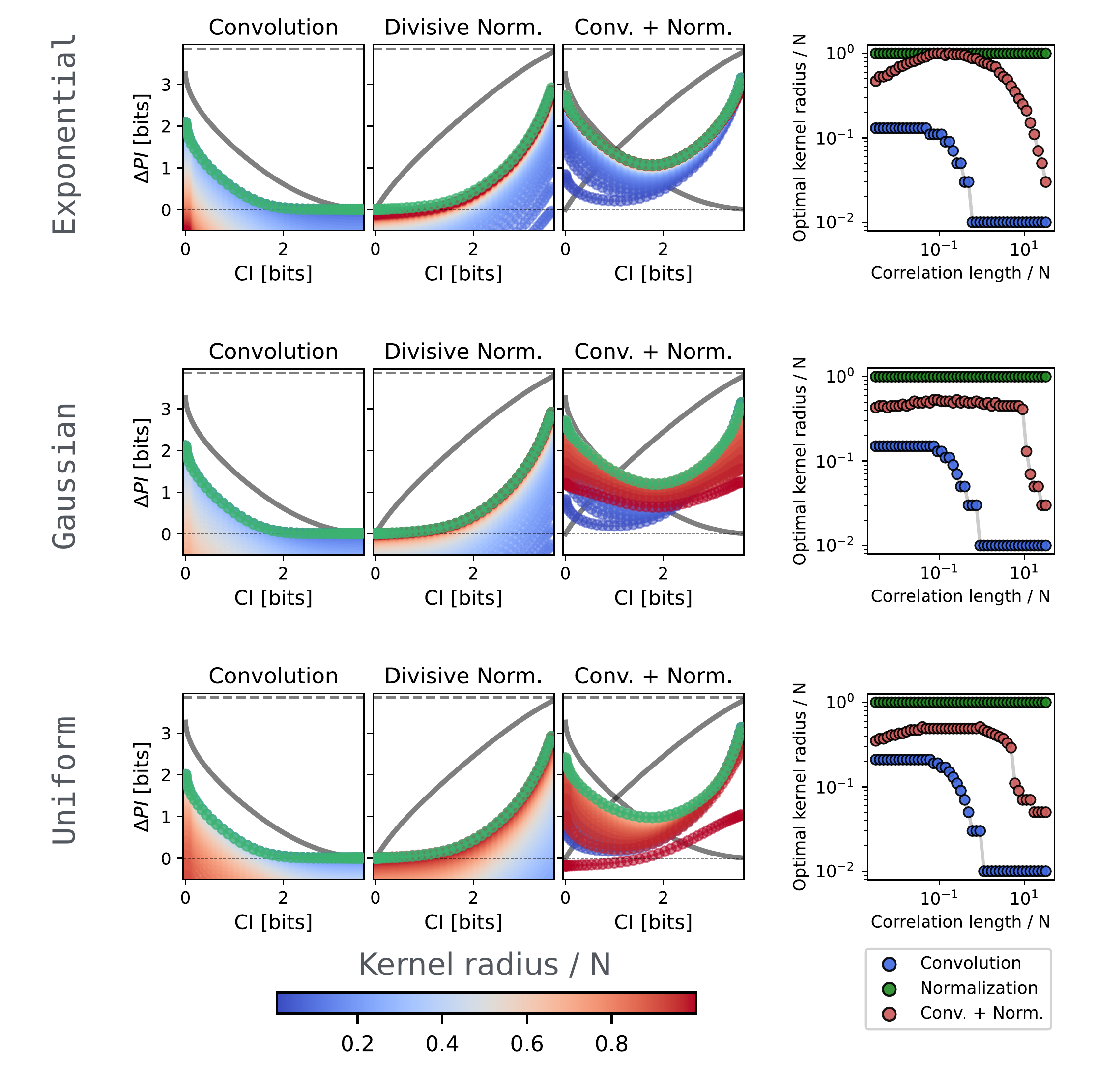}
    \caption{Algorithmic $\Delta PI$ for convolution, divisive normalization, and their sequential application. The three rows correspond to different shapes of the convolution kernels. The color map represents the size of the convolution kernel. Right column: optimal kernel sizes.}
    \label{fig:different_kernels}
\end{figure}

We also tested alternative ways of computing the normalization constant in the denominator, using different weighting kernels (uniform, Gaussian, and exponential) and varying the spatial range over which the normalization is performed. In all cases, the best performance is consistently obtained when the normalization is computed over the entire embryo. (See Figure~\ref{fig:different_kernels})

Taken altogether, these results show that ALP decoding reduces to divisive normalization in the limit where $L_{corr} \rightarrow \infty$. If the local variance modulation were instead spatially uniform, $v(x) = 1$, the same limit would yield a form of ``{\it normalization by subtraction}", rather than by division (Figure~\ref{fig:other_approx_ALP}(a)).

\subsection{More complex algorithmic approximations}\label{sec:other_ALP_RLP}
In the previous sections, we showed that optimal nonlocal decoding under RLP and ALP admits simple algorithmic interpretations in limiting regimes: local decoding after spatial smoothing for RLP with weak correlations, and divisive normalization for ALP with long-range correlations. Here, we extend this analysis by exploring more general algorithmic approximations that approach the optimal theoretical bounds more closely.

Our strategy is to consider linear transformations of the morphogen profiles,
\begin{equation}
    \mathbf{g}^{\rm eff} = \mathbf{L}\mathbf{g}
\end{equation}
followed by standard local decoding applied to the transformed signal. We then optimize the transformations to maximize positional information.

\subsubsection{Approximations for Relative Locations Prior (RLP)}
In Section~3~A, we found that RLP decoding can be approximated by local decoding from an effective profile $\mathbf{g}^{\rm eff}$ obtained via spatial averaging. While this simple convolutional scheme is easy to interpret and implement, there could be more general linear operators $\mathbf{L}$ which could further increase positional information.

We first explore spatially varying exponential kernels of the form
\begin{equation}
    L_{ij} \propto \exp(-|i-j|/ \ell_i) 
\end{equation}
where the length scales $\ell_i$ may be cell-specific. Standard convolutions correspond to the special case $\ell_i \equiv \ell$. By optimizing the set of $\ell_i$, we maximize the positional information of the transformed profiles $\mathbf{g}^{\rm eff}$.

As a second variant, we allow for asymmetric weighting of neighbors to the left and right of each cell:
\begin{eqnarray}
    L^{{\rm asym}}_{ij} &\propto& \exp(-|i-j|/\ell^{{\rm left}}_i) \qquad \,\,\, {\rm if} \,\,\,j < i \\
    L^{{\rm asym}}_{ij} &\propto& \exp(-|i-j|/\ell^{{\rm right}}_i) \qquad {\rm if} \,\,\,j > i
\end{eqnarray}
This relaxation captures directional biases in information integration, which may arise from asymmetries in the morphogen profile of noise structure.

Finally, we consider a third construction derived directly from the likelihood maximization problem underlying optimal RLP decoding (Equation~\ref{eq:RLP_likelihood}). 

For the specific case of a simple exponential mean profile $\bar{g}(x)= \mu \exp(-x/\lambda)$ the following holds:
\begin{eqnarray}
    \bar{g}(x_i^* + d_{ij}) &=& \bar{g}(x^*_i)e^{-d_{ij}/\lambda} \\
    \bar{g}'(x_i^* + d_{ij}) &=& -\bar{g}(x_i^*)e^{-d_{ij}/\lambda}/\lambda
\end{eqnarray}
Substituting these relationships in Eq.~\ref{eq:RLP_likelihood} and neglecting the spatial variation in the noise amplitude, the optimization problem then becomes:
\begin{align}
    \frac{dL(x_i^{\ast})}{dx_i^{\ast}} & = \sum_j \bar{g}'(x_i^{\ast} + d_{ij})\left[ \sum_k C^{-1}_{jk}\big( g_k - \bar{g}(x_i^{\ast} + d_{ik}) \big) \right] =  \\
    & = -\sum_j \frac{\bar{g}(x_i^{\ast})e^{-d_{ij}/ \lambda}}{\lambda} \left[ \sum_k C^{-1}_{jk} \big( g_k - \bar{g}(x_i^{\ast}) e^{-d_{ik}/\lambda} \big)\right] = 0
\end{align}
This equation is solved by $\hat{x}_i$ such that:
\begin{equation}
    \bar{g}(\hat{x}_i) = \sum_k g_k L_{ik} \equiv g^{\rm eff}_i
\end{equation}
with 
\begin{equation}
    L_{ik} = \frac{\sum_j e^{-d_{ij}/\lambda}C^{-1}_{jk}}{\sum_{j,m}e^{-(d_{ij}+d_{im})/\lambda}C^{-1}_{jm}} 
\end{equation}

\begin{figure}
    \centering
    \includegraphics[width=0.85\linewidth]{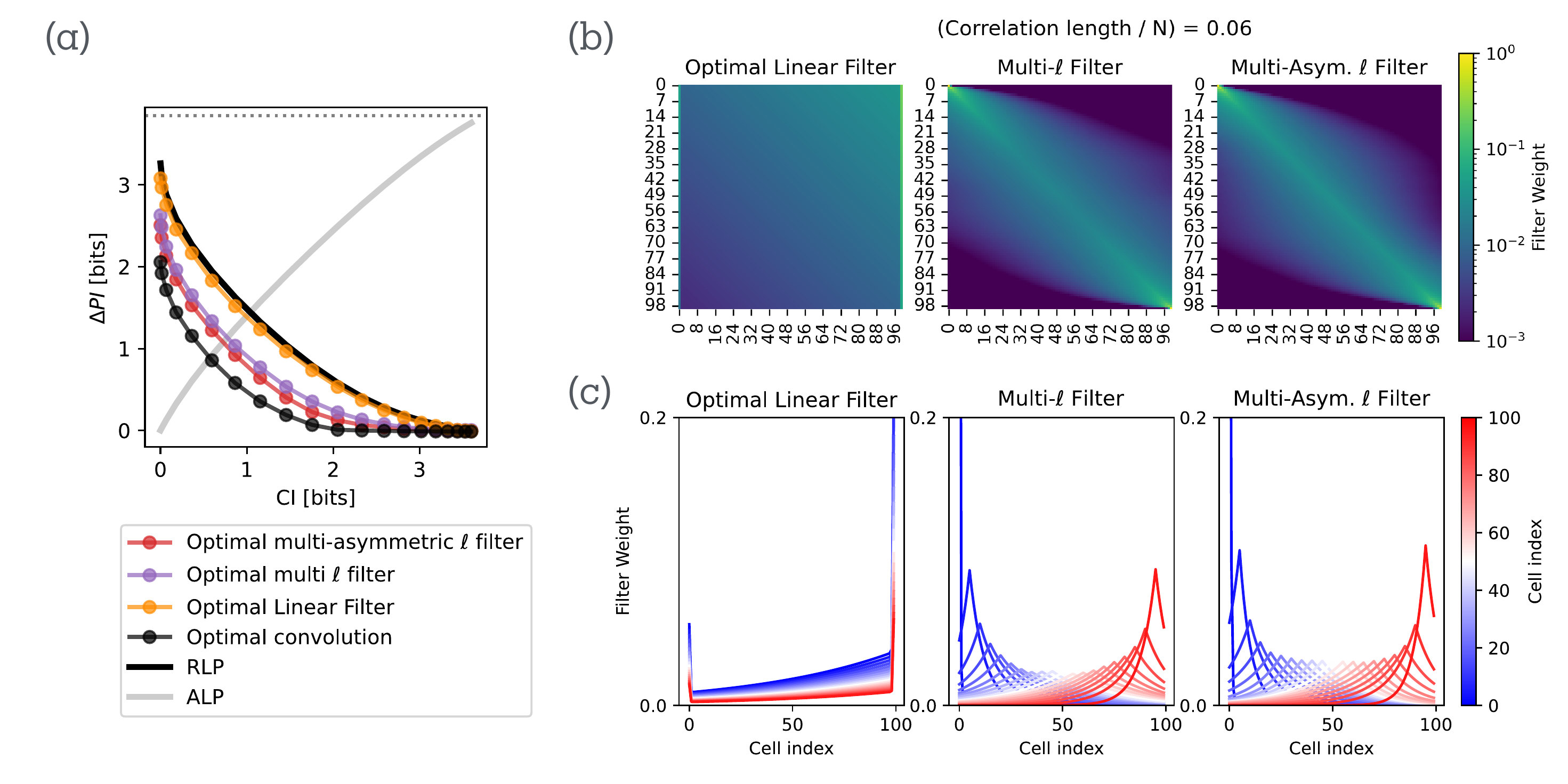}
    \caption{
    {\bf (a)} Gain in positional information (PI) obtained from the different linear transformations $\mathbf{L}$ introduced in Sec.~3~C. The orange curve saturates the optimal RLP bound but corresponds to a transformation that is less readily interpretable than the convolution-like approximations (red and purple). All three constructions achieve performance intermediate between simple convolution and the optimal RLP limit. 
    {\bf (b)} Heatmaps of the corresponding transformation matrices $\mathbf{L}$ for the three cases. 
    {\bf (c)} Same data as in (b), shown row by row: each curve corresponds to one row of $\mathbf{L}$, with color indicating cell index. For the convolution-like transformations, the kernels are centered near the target cell, reflecting a degree of spatial locality. In contrast, the optimal-performing transformation exhibits less localized and harder-to-interpret structure, making a direct mechanistic implementation less apparent.
    }
    \label{fig:other_approx_RLP}
\end{figure}

For all three constructions, we compute the resulting positional information gain from the effective profiles $\mathbf{g}^{\rm eff}$. As shown in Figure~\ref{fig:other_approx_RLP}, each refinement improves performance relative to simple convolution, with the covariance-informed filter fully saturating the optimal RLP bound. These results demonstrate that optimal RLP decoding can be achieved algorithmically through linear transformations, albeit at the cost of reduced interpretability.

\subsubsection{Approximations for Absolute Locations Prior (ALP)}

We now turn to an additional algorithmic approximation of ALP decoding. Recall that under ALP, the optimal estimate of position is obtained by matching the observed morphogen level $g_i$ to the conditional mean,
\begin{equation}
    g_i = \tilde{g}(\hat{x}_i) = \bar{g}(\hat{x}_i) + C_{i, \neg i}[\mathbf{C}_{\neg i, \neg i}]^{-1}(\mathbf{g}_{\neg i} - \bar{\mathbf{g}}(\mathbf{x}_{\neg i})
\end{equation}
Rearranging, this implies an effective profile
\begin{equation}
    \bar{g}(\hat{x}_i) = g_i - C_{i, \neg i}[\mathbf{C}_{\neg i, \neg i}]^{-1}(\mathbf{g}_{\neg i} - \bar{\mathbf{g}}(\mathbf{x}_{\neg i}) \equiv g_i^{\rm eff}
\end{equation}
from which one can perform local decoding.

To assess whether this can be approximated algorithmically, we seek a linear mapping $\mathbf{L}$ such that:
\begin{equation}
    \mathbf{g}^{\mathbf{\rm eff}} = \mathbf{L}\mathbf{g} + \mathbf{\epsilon}
\end{equation}
where $\mathbf{\epsilon}$ is an error vector. Using an ensemble of morphogen profiles collected across $M$ embryos, we estimate $\mathbf{L}$ by minimizing the Frobenius norm of the residual error:
\begin{equation}
    \hat{\mathbf{L}} = \rm{argmin}_\mathbf{L}||\mathbf{G}^{\rm eff} - \mathbf{L}\mathbf{G}||
\end{equation}

\begin{figure}
    \centering
    \includegraphics[width=0.85\linewidth]{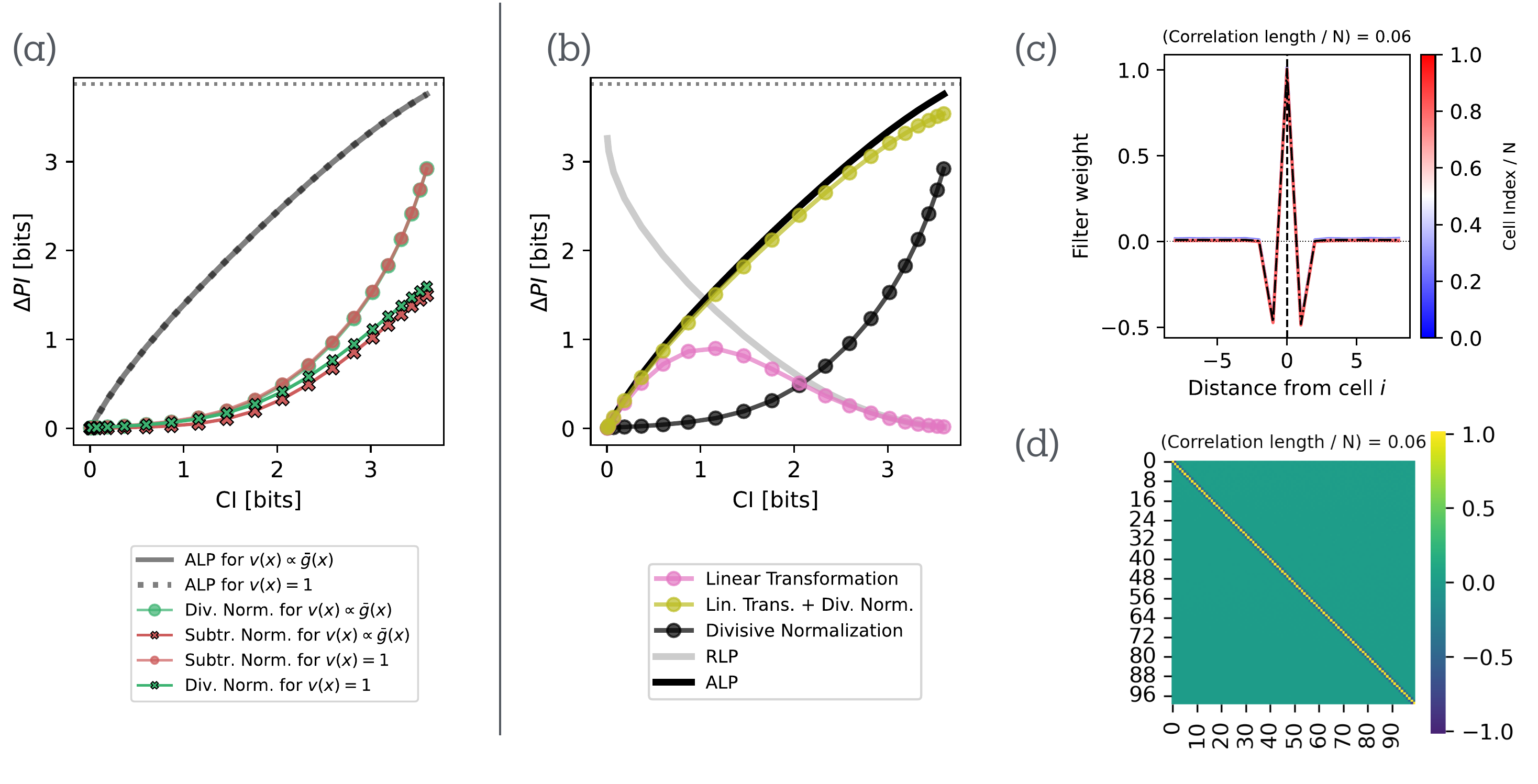}
    \caption{
    {\bf (a)} Gain in positional information, $\Delta \mathrm{PI}$, achieved by divisive versus subtractive normalization for morphogen ensembles with spatially uniform noise amplitude ($v(x)=1$) or spatially varying amplitude ($v(x)\propto \bar{g}(x)$). As predicted by the theory, divisive normalization performs best when noise scales with the mean profile, while subtractive normalization is optimal for spatially uniform noise. 
    {\bf (b)} Algorithmic approximation to ALP decoding discussed in Sec.~3~C. A purely linear transformation $\mathbf{L}$ is insufficient to approach the ALP bound, whereas combining linear preprocessing with normalization effectively suppresses both short- and long-range correlated fluctuations. 
    {\bf (c,d)} Example structure of the learned linear transformation $\mathbf{L}$ for a representative correlation length.
    }
    \label{fig:other_approx_ALP}
\end{figure}

Applying this linear transformation and decoding locally from $\mathbf{g}^{\mathbf{\rm linear}} = \hat{\mathbf{L}}\mathbf{g}$, we find that purely linear transformations fail to reproduce the ALP information gain (Figure~\ref{fig:other_approx_ALP}(b)).
However, when this linear processing is followed by a nonlinear divisive normalization step, performance nearly saturates the optimal ALP bound.

\section{Reaction diffusion model: steady state solution and simulation details}\label{sec:reaction_diffusion}
Here, we provide additional details on the reaction-diffusion network introduced in the main text as a minimal biochemical implementation of the algorithmic approximations to the optimal RLP and ALP decoding. 

As described in the main text, we consider the following reaction-diffusion system:
\begin{eqnarray}
\frac{\partial n_{1}}{\partial t} &=& D_1 \nabla^2 n_1 + \alpha_1 g - \kappa_1 n_1 \nonumber \\[2mm] 
\frac{\partial n_{2}}{\partial t} &=& D_2 \nabla^2 n_2 + \alpha_2 n_1 - \kappa_2 n_2 \label{eq:rd_system} \\[2mm]  
\frac{d g^{\rm out}}{d t} &=& \alpha^{\rm out} n_1 - \kappa^{\rm out} n_2 g^{\rm out}, \nonumber
\end{eqnarray}
with $D_2 \gg D_1$.

The input morphogen profile $g(x)$ is taken to be static on the timescale of the dynamics, i.e. $g(x,t) = g(x,0)$. The fields $n_1$ and $n_2$ represent intermediate biochemical species that process the input profile, while $g^{\rm out}$ denotes the final processed signal.

\subsection{Steady-state solution and interpretation}
We begin by analysing the steady state solutions of Eqs.~\ref{eq:rd_system}. Setting $\partial_tn_1=0$, the equation for $n_1$ becomes:
\begin{equation}
    \left( \nabla^2 - \frac{1}{\ell_1^2} \right)n_1 = -\frac{\alpha_1}{D_1}g(x)
\end{equation}
where $\ell_1 = \sqrt{\kappa_1/D_1}$ defines the characteristic diffusion length of $n_1$. This is the screened Poisson (Helmholtz) equation. Its solution can be expressed in terms of the Green's function $G(x,x')$ of the linear operator $(\nabla^2 - 1/\ell_1^2)$, defined by:
\begin{equation}
    \left( \nabla^2 - \frac{1}{\ell_1^2} \right)G(x,x') = -\delta(x,x')
\end{equation}
The steady-state profile $n_1(x)$ is then:
\begin{equation}
    n_1(x) = \frac{\alpha_1}{D_1}\int G(x,x')g(x')dx'
\end{equation}

For a one-dimensional infinite domain--an accurate approximation away from boundaries when $\ell_1$ is smaller than the system size--the Green's function takes the form
\begin{align}
    G(x,x') = \frac{\ell_1}{2}\exp\Bigg(-\frac{|x-x'|}{\ell_1}\Bigg)
\end{align}
Thus, $n_1(x)$ is given by a convolution of the input morphogen profile with an exponential kernel of range $\ell_1$. At steady state, the first layer of the network therefore implements spatial smoothing of the input signal, consistent with the convolutional preprocessing identified in the algorithmic approximations to RLP decoding.

The second species $n_2$ obeys an analogous equation but takes $n_1$ as its input. In the limit $D_2 \gg D_1$, diffusion dominates the dynamics of $n_2$, and its steady-state profile becomes approximately spatially uniform. In this regime, $n_2$ is proportional to the spatial average of $n_1$, and thus to the global average of the input morphogen profile. The species $n_2$ therefore acts as a global integrator of morphogen levels across the embryo.

\begin{figure}
    \centering
    \includegraphics[width=0.99\linewidth]{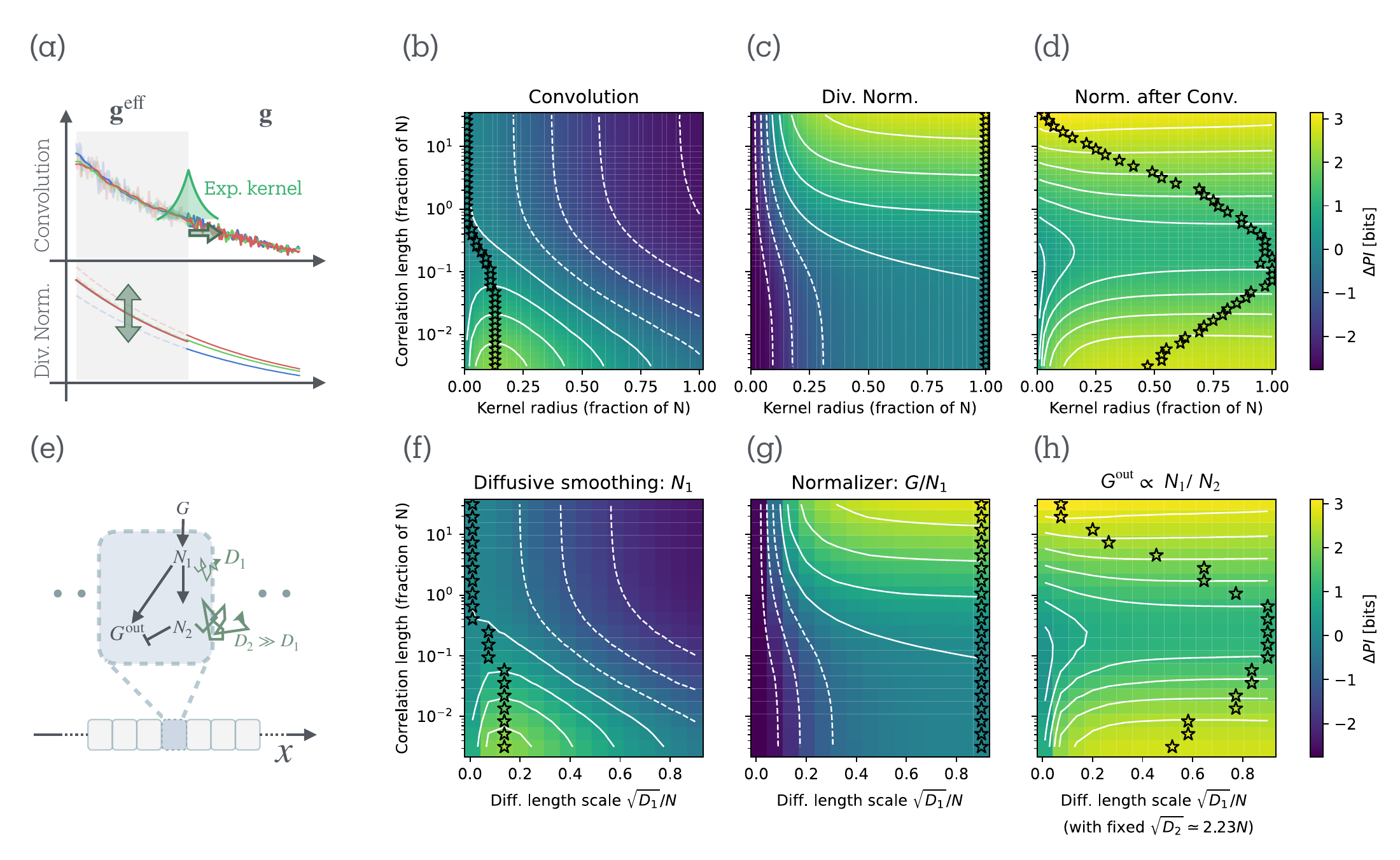}
    \caption{{\bf (a)} Algorithmic transformations map morphogen profiles $\mathbf{g}$ into effective profiles $\mathbf{g}^{\rm eff}$ via spatial convolution and/or divisive normalization; these effective profiles can be subsequently decoded locally and optimally. $\Delta \mathrm{PI}$ achieved by: {\bf (b)} convolution with exponential kernels of varying radii across a range of $L_{\rm corr}$ values; {\bf (c)} divisive normalization with a normalization constant obtained via convolution with kernels of varying radii; {\bf (d)} sequential application of convolution (variable radius) and global normalization (sum over all $N$ cells). {\bf (e)} A minimal reaction-diffusion circuit with two diffusible {\it normalizer} species ($N_1$, $N_2$) implements algorithmic operations of (a-d).
    {\bf (f-h)} Same analyses as in {\bf (b-d)}, but using the reaction--diffusion implementation. 
    {\bf (f)} $\Delta \mathrm{PI}$ associated with the steady-state profile $N_1$ while varying its diffusion length $\sqrt{D_1}/N$. 
    {\bf (g)} $\Delta \mathrm{PI}$ associated with $G/N_1$ while varying $\sqrt{D_1}/N$. 
    {\bf (h)} $\Delta \mathrm{PI}$ associated with $N_1/N_1$ while varying $\sqrt{D_1}/N$, with $\sqrt{D_2}/N$ fixed at $\approx 2.23 \gg \sqrt{D_1}/N$. 
    In (b-d) and (f-h), stars identify the optimal kernel radius for each correlation length value.
}
    \label{fig:SM_algos_and_mechanism}
\end{figure}

Finally, the output signal $g^{\rm out}$ is produced by $n_1$ and degraded at a rate proportional to $n_2$. At steady state, Eq.~\ref{eq:rd_system} implies
\begin{equation}
    g^{\rm out}(x) \propto \frac{n_1(x)}{n_2(x)}
\end{equation}
The last layer of the network thus performs spatial convolution followed by divisive normalization. As shown in Fig.~\ref{fig:SM_algos_and_mechanism}, the positional information gains achieved by this minimal biochemical circuit quantitatively match those obtained from the corresponding algorithmic approximations.

\subsection{simulation details}
All simulations were performed in one spatial dimension on a domain $x\in [0, L]$, discretized into $N$ sites with $L=N$ and lattice spacing $\delta x = 1$. The input morphogen profile was fixed in time, $g(x,t)=g(x,0)$, and the initial conditions for the internal species were $n_1(x,0)=n_2(x,0)=0$. 

Time integration was carried out using an explicit Euler scheme with timestep $\delta t = 10^{-5}$. Spatial second derivatives were approximated using centered finite differences. We used von Neumann (closed) boundary conditions, enforcing vanishing flux at the boundaries of the simulation domain. 

Parameters were chosen as:
\begin{equation*}
    \alpha_1 = \alpha_2 = \alpha^{\rm out} = \kappa_1 = \kappa_2 = \kappa^{\rm out} = 1, \qquad D_2 = 5 \times 10^{4}, \qquad D_1 \in [1, 8100] 
\end{equation*}

For each parameter set, simulations were run until steady-state was reached, and the resulting output profiles were used to compute positional information as described in the main text.

\clearpage

\section*{Supplemental Figures}\label{app:figs}

\clearpage

\clearpage


\clearpage


\onecolumngrid

\end{document}